TARTU ÜLIKOOL
Majandusteaduskond
Juhtimise ja välismajanduse instituut
Välismajanduse õppetool

Jaan Übi

**TIPPJUHI ROLL SISEMISE KONKUREERIMISE JA TEADMISTE VABA JAGAMISE VAHELISE TASAKAALU LEIDMISEL ALLÜKSUSES**

**GENERAL MANAGER'S ROLE IN BALANCING SUBSIDIARY BETWEEN INTERNAL COMPETITION AND KNOWLEDGE SHARING**

Bakalaureusetöö

Juhendaja: Urmas Varblane, Zuhair Al-Obaidi

Tartu 2003

# TABLE OF CONTENTS

# INTRODUCTION

Today multinational corporations (MNC) are, in order to face diverse demands, organized in more complex ways than ever before. In the reorganizations their local country operations – subsidiaries – have been transformed.

From the beginning of 1980s there is a distinguishable stream of research in MNC theory that focuses on subsidiary issues. This research has risen mainly from countries, which consider MNC subsidiaries important for their economies, such as Canada, UK, Spain, Denmark and Sweden, and deals with issues such as headquarters-subsidiary relationships, subsidiary strategies and subsidiary development.

There are a couple of reasons, why such explicit focus on subsidiary issues is all but natural. Firstly, it is the position of former, independent 'miniature replicas' that can considerably deteriorate in the rationalization-integration into MNC's global operations. Subsidiaries have developed capabilities and would like to be more than just parts of a system controlled from outside (Morrisson & Roth [1993→]).

Secondly, it is a fact, that in spite of all the process of globalization, there still is a strong justification for subsidiaries as such. There are reasons on 'demand side' – as differing wishes give inspiration to local units – and on 'supply side'.

On 'supply side', because despite the operations being increasingly footloose, there still are 'sticky' capabilities, that are hard to transfer (Birkinshaw & Hood [1998b→]). Also because despite the increasing mobility of people, most of them will still spend the main part of their career in home country; and despite of developing information technology, there are still advantages to geographical proximity.

Current thesis is also focusing on MNC subsidiaries. It is in line with works of Birkinshaw [1995a→, 2000→], which claim that in the framework of local responsiveness and global integration (Prahalad & Doz [1987→]), MNCs are perhaps currently relying too much on product divisions (global integration). So, for instance, in order to take different initiatives within MNCs, subsidiaries must become more "stocky" – not so lean and tightly integrated.

Current thesis does not speak of the geographical dimension, though. It explicitly considers the uppermost position on that dimension – the subsidiary general manager (SGM) – and brings out contributions this post can make. This way it leaves out cases such as advocating for the whole dimension and reorganizing through uniting all marketing forces within a country (in order to gain customer servicing related synergies→ as discussed in Forsgren & Johansson [1992→]).

The point of origin of the thesis can be said to be in the statement made by Birkinshaw [1995a→]: '*there is a need for strong subsidiary general managers to provide coherent vision for subsidiary /…/ So that the subsidiary would not just be a group of value-added activities that happen to be located in one country'. As opposed to statements made by product dimension proponents – Humes [1993→] claiming that country managers should be weakened at their position, stating emotionally 'there is a need to lessen the degree of influence of country barons'. Or from anecdotal evidence, a case study by Hedlund & Ridderstrale [1995→] finding instances of bounded rationality in rationalization-integration processes of MNCs in the case that has strong subsidiary top management, but not in the case where product divisions dominate.

The study sets out to explore this contradiction in literature, to find out *how authority position helps subsidiary general manager to contribute in a MNC*. The issue is viewed in light of recent conceptualizations of MNCs that leave space for further investigation.

From one side literature has emphasized internal competition within MNCs, resulting in, for example, its definition through internal markets (Birkinshaw [2000→]). At the same time focus has been on achieving maximal leverage of MNCs' capabilities through effective knowledge sharing and cooperation (Ghoshal & Bartlett [1997→]).

As MNCs have to find a delicate balance between these countering requirements, the purpose of this study is to explore *what is the role of subsidiary general manager in* *balancing subsidiary between internal competition and centre of excellence type of knowledge sharing.*

This work is consists of an analysis of literature and empirical research the former provides a basis for. It seeks answers to the following research questions:

- What are the relevant characteristics while considering MNCs generally, and for the SGM post in particular?

- How have recent theoretical developments emphasized MNC's multi-centeredeness, thereby giving subsidiaries an increasing attention?

- How can we classify subsidiaries for the purpose of selection and differentiation?

- How does the process of internal competition differ from a focus on knowledge-sharing, and what is the role of subsidiary general manager in both cases?

The research problem investigated has so far not been described in necessary depth and thus calls for qualitative exploratory research. The empirical research conducted in this thesis is carried out in form of case studies by conducting interviews.

The companies studied are two Estonian subsidiaries of big multinational corporations, both being worldwide leaders in their respective business areas. Both subsidiaries in turn meet the condition of being 'substantial' enough for it to be possible to apply concepts discussed in contemporary MNC literature.

The thesis is organized in three chapters. The first two outlining theoretical basis and the third presenting results of empirical investigation. The first chapter takes the level of the multinational as a whole. It speaks of MNCs' characteristics in 1.1. and development of theories in 1.2. Also throughout the discussion additional points concerning SGM's post are brought up. In the end of the chapter, characteristics are linked to the SGM's discussion.

Second chapter gives a classification of subsidiaries in 2.1. In 2.2 the process of internal competition is outlined. The knowledge sharing processes are discussed in 2.3. In the end a hypothesis is put forth for empirical investigation.[i]

In 3.1 of the third chapter an overview is given of companies studied showing characteristics that are of further relevance for the study. Part 3.2 presents the results of case study interviews. In the end the work is concluded with a discussion. Further implications for research are also outlined.

It is interesting to note that Estonian research into MNC subsidiary management was practically non-existent at the time, when the first articles used in this work were gathered. This although foreign investments have been very important for Estonia, and operations of MNCs subsidiaries have accounted for many successful cases in development of the economy. Hopefully this work has also helped by discussing some articles that are worth further investigation.

This work would not have been possible to write without the help and co-operation from Estonian subsidiaries ES Sadolin and Saint-Gobain Sekurit Eesti. Author is deeply indebted to both the organizations and individuals for taking their time to participate in the study and co-operating in a splendid way. Author is also very grateful to his tutors Urmas Varblane and Zuhair Al-Obaidi from Helsinki School of Economics for their guidance and advice. Author would also like to thank Julian Birkinshaw from London Business School for his answers concerning specific topics of subsidiary management.

# 1. HETERARCHICAL MNCS

## 1.1. Characteristics of MNCs resulting from changing environment

During the 20$^{th}$ century Multinational Corporations (MNC) grew enormously both in size and number and, in parallel with current expansion of capital sphere vis-à-vis social sphere(Ruigrok & van Tulder [1995→]), are taking on an increasing role in world affairs. In settings which are (acceleratingly) changing the MNCs have themselves retooled. As an example of their influence on theory Hedlund [1993→], pointing to one work by Casson, writes that 'some lines of research take the view that MNC should be the general case for the theory of the firm to discuss, so that simpler types of company would be treated as special cases'.

To start analyzing the MNCs, it is important to show some of their characteristics. It is the point of view of this work that MNCs should be considered as *essentially three-dimensional*. An MNC can thus be characterized by its *product* (eg. products h1, h2, h3, i1, i2 and divisions H, I$^{ii}$), *geographical*, and *functional* properties (eg. marketing, production and R&D).

Humes [1993→] brings examples from world's biggest companies throughout the 20$^{th}$ century that have in the past shown themselves to be archetypal, built dominantly around any one of those dimensions. Perlmutter's [*1969→] polycentric MNC is the one built around geographical lines and geocentric is a product (division) company. Many Japanese MNCs have been documented at using function based structures (Humes [1993→]).

We are now going to discuss environmental changes$^{«}$ that have taken place recently and had their influence on the way MNCs organize along the three dimensions. As discussed for example in Buckley & Casson [1998→], UNCTC [1993→], Ghoshal & Bartlett [1995→], Birkinshaw [1996→], Porter [1986→] and Jeannet[2000→], by the middle of 1970s trade liberalization was well underway.

One big consequence for MNCs was the process of rationalization – from tariff barrier separated geographical structure to product (division) structure. We will come back to this issue in section 2.1[iii] and for now it is enough to say that MNCs tended to organize around product divisions.

The middle 1970s were the end of the postwar Golden Era of growth, in which economies had been moving quickly and unidirectionally upwards. As economic climate changed it could thereafter best be characterized as less pacedly growing and increasingly* volatile.

As for lower growth oil-crisis was one factor. In part it has also been attributed to economic policies that concentrated on control of inflation and not on expansion of demand, and to the fact that major industrial countries failed to co-ordinate their policies so as to ensure steady growth combined with internal and external equilibrium.

In the following period trade liberalization process was to some extent retracted – through raising non-tariff barriers. And as Bretton Woods system collapsed flexible exchange rates were put into use. Up to that point the progressive (predictable) dismantling of trade barriers and highly estimable exchange rates between major trading nations had arguebly been enabling the high growth of the Golden Era.

Flexible exchange rates were one emergent source of volatility. Another was that, as a consequence of liberalization processes, world economy had become increasingly interdependent and thus the shocks in it spread more widely. The interest rates for instruments in use within countries became more volatile. As for the fluctuation of economic growth itself, it can be said that the two biggest recessions since 1930s happened in the next decades. As the Far-Eastern economies rose another source of volatility was added – the interdependent world economic system was thereafter simply determined by more variables.

Some more things should be noted about the emergent situation. As a bigger number of economic powers were relevant in world's economic system, there were now more regions to be drawn upon for their capabilities, that is through subsidiary corporations. Emerging Far-Eastern corporations meant increasing competition for Western companies.

Moreover, competition intensified in growing globalization of demand, and also due to the low growth climate and expanding R&D costs. As for example foreign direct investments (FDI) into US had been insubstantial, it increased considerably indicating among other things a battle for US MNCs home turf. Lastly it should be noted, that if environment of MNCs subsequently contained more information, the developments in information technology field enabled its more efficient dispersal in compensation.

Conceiving the resultant, much more complex situation it became clear that 'MNCs could no longer exist the way they had been [during the Golden Age] – as top-down-systems driven formations, where lower levels were for the purpose of gathering information, handing it upwards first and acting on the intention of the brain of the firm later' (Ghoshal & Bartlett [1995→]). Also studies that investigated MNCs found them to be structurally very complex embodiments, that in our three-dimensionality discussion can be labeled to have *a divergent, partially overlapping structural map*.

In order to cope with the volatile environment MNCs drive for flexibility. They are flexible by using (ever-changing) heterogenous control relations – that is there are different centers in MNCs that have different areas of control.

As displayed in Figure 1 a, say Australian, subsidiary of an MNC may have developed into World Product Mandate (WPM – cf. Roth & Morrison [1992→]) assignee. This means that it has acquired control of the full value-added scope (eg. logistics, R&D, production and marketing) of a specific product or product line (a product from the chemicals division – $c_1$) with responsibility of producing for world market and controlling all sales operations.

On the other hand the sales people in Stockholm (thus the Swedish subsidiary) are so successful in creating home-electronics division's (H) marketing campaigns, that they are formally acknowledged as its marketing Centre of Excellence (CoE – cf. Moore & Birkinshaw [1998→], more closely discussed in 2.3[iv]) for the Western hemisphere*.

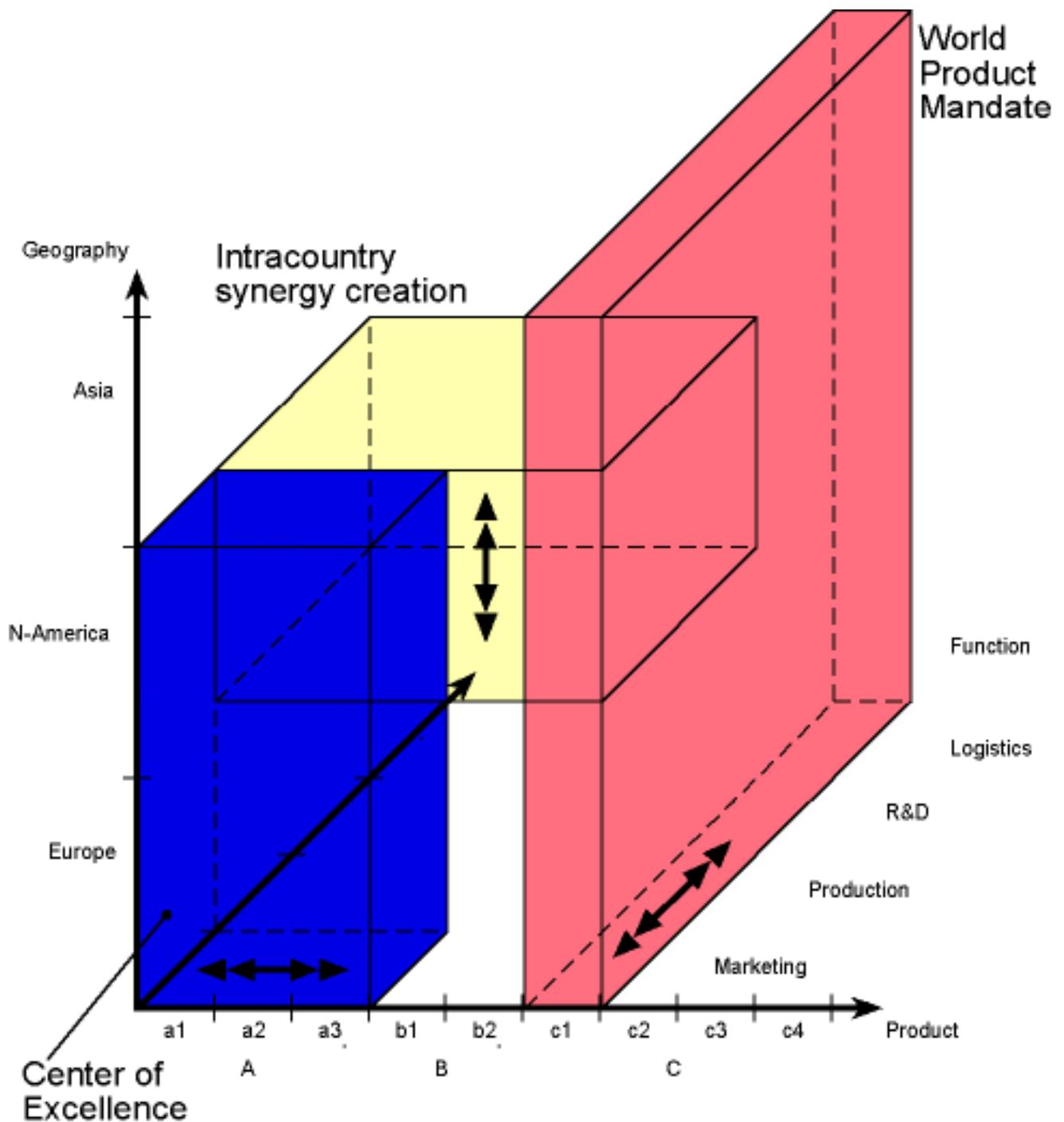

Figure. A divergent non-overlapping structural map of an MNC

The third example would be a case of facilities of home and industrial electronics divisions (H, I) in the US. As there is a big volume of operations in that subsidiary, it is spotted that the cross use of plants between the products can be of a benefit. Additionally synergies are sought for in R&D as the technological fields are related and scientific personnel tend to work in national research clusters[vi]. (For example Birkinshaw [1995a] mentions complementary activities in a subsidiary).

Thus if in WPM control was exercised on product dimension and CoE exhibited function based arrangement, then here the coordinating, transporting and communicating work is feasibly intensified within US and geographical dimension prevails.

Important issues for this work are the emerging *conflicts of interest,* as already the main research issue – competing and sharing – is a conflict situation. But hereby we are going to begin with a basic case of disagreement between R&D and marketing. R&D department prefers little communication as marketing is said to be giving away all the secrets to competitors, whilst campaigning; and marketing blames the lack of communication for the creation of supposedly out of touch products, that do not succeed in the marketplace.(Hedlund & Ridderstrale [1995])

Another case of conflict can be constructed in our three-dimensional space. Let the intracountry synergy creation this time consider marketing. Some works from interorganizational theories school consider development of relationships in business networks (eg. Forsgren & Pedersen [1998], Forsgren & Johansson [1992], Forsgren & Pahlberg [1992]). They state that due to mutual adaption in practice business relationships often develop into something between the arms length market interaction and strict hierarchical 'fiat'.

As it was also mentioned in introduction, Forsgren & Johansson [1992] write subsequently that in MNCs there is an effect in combining marketing operations with often shared customer bases as is in our case of home and industrial electronics division. Now lets say that there is a home electronics product, cassette player, that is WPMandated to Taiwan's subsidiary. This means that in addition to producing it,

Taiwan should also control the marketing operations worldwide. And thereby we see a conflict between two simultaneous goals.

As a possible way of resolving such a conflict, which by the way can be said to take place between two subsidiaries[vii], both trying to enhance their sphere of influence[→]; Forsgren & Johansson [1992[→]] write that marketing units can be combined into profit centers with some control (such as over staffing decisions) being given outside. We can however conclude, that if Figure 1 showed a divergent structural map of an MNC, then the current example represents a structure with divergent and partially overlapping control relations.[1]

## 1.2. Latest development of organizational aspects in MNC theories

One important step in development of MNC theory have been the seminal works of Bartlett and Ghoshal, in which the concept of Transnational was developed. Transnational (cd.[2] eg. Bartlett & Ghoshal [1992[→]]), balancing effectively the differing pressures MNCs face, is a company operating a divergent, partially overlapping map that we previously described.[viii]

It is, however, interesting to look at the way the structure of Transnational is presented. Authors [1992[→]] say that companies have one general macrostructure, for example a product division structure. The macrostructure is complemented by relationship between line (of command) and staff of a MNC.[ix] And the complex structural demands are solved with microstructural tools − elements of microstructure are besides different control relations we described, for example, task forces and committees. Authors also coined the term administrative heritage of corporation and, for example, in Ghoshal and

---

[1] At this point we can also say, why MNCs should be considered essentially three-dimensional. Because this way we can see different processes as having a common origin, only applied along varying lines. For instance, if CoE is basically a mean for leveraging information across geographical boundaries, then intracountry synergy creation does the leveraging across divisional boundaries. And the case that had coordinating problems between R&D and marketing, being an example of interdependencies and a need for integration, can also be said to need leverage across, this time, functional boundaries.

[2] Cd. is used in this work as an abbreviation marking that Concept (is) Discussed in a certain work. It is applied to simplify distinguishing cases where the concept used in sentence is investigated in another work, in general, form and cases where current sentence quotes a specific idea put forth by another author.

Bartlett [1990→] say that this might have a great deal of influence on which actual macrostructure is used in a corporation.

Thereby if once the "structural archetype" that was used in MNC might have been a very good descriptor, now the fact that, for example US, European and Japanese companies tend to use different macrostructures is no longer as decisive. These are the lower level structural solutions that are similar, and that matter. This is actually a step forth from the work of Humes [1993→], who still emphasizes archetypes and operates with them.

Later works on MNCs have gone still further as they have arrived at a question whether it is at all feasible to differentiate the uppermost (or overriding) hierarchy in organizations. The works of Hedlund [1986→, 1990(with Rolander)→, 1993→] draw on manifests for change of perspectives in a wide range of sciences. The main idea being that the reality is actually organized non-hierarchically and that we are only accustomed to working with it through hierarchies. One good example from complex embodiments is highest level in evolution – the brain (or neural network) –, functioning of which we can not completely explain, but which surely is a non hierarchical system.

Hedlund [1986→] speaks of MNCs as heterarchical and describes their different characteristics. In heterarchy works, organizations also have a partially overlapping and divergent structural map. But here the units are termed centers, each having a different degree of influence – which is, for example, one of the validations for the claim that no structural dimension should be considered superordinate to the rest. In accord with stripping the corporation of the macrostructural tool of transnational, Hedlund [1993→] brings examples of corporations that take pride for not having an organizational chart published at all.

There are also other works with similar point of view. The works on business networks(eg. Forsgren & Pedersen [1998→]) often do not make qualitative differentiation between cases wherein an organizational unit is involved in transactions with sister units and cases wherein transactions take place with external parties. Thus we can again speak of *multicentered view of MNC*. Some business network articles have also modeled the positions of subsidiaries (centers) within MNCs in terms of influence (Forsgren & Pahlberg [1992], also Ghoshal & Bartlett [1990])[3].

But the works of Ghoshal and Bartlett ([1995], [1997], [1999](with Moran)) also take a new path. In their new book on individualized corporations [1997] the emphasis is put on changing the main terms used while considering a company. They conclude the past paradigm to have been Strategy-Structure-Systems and put forth a new one – Purpose-Processes-People. Thereby the hierarchical (macro)structure is left out of their focus too, although the accent is a bit different from Hedlund's works (eg. Hedlund & Rolander [1990], where heterarchical (structure) was the base according to which the company was handled).

The works of Burgelman[x] [1983, 1994] that consider autonomous strategic decisions in big corporations are also relevant in this context. These works show that multiple layers of management are actually involved in taking strategic decisions. This way the managers, who are at first to be considered lower at the organizational drawing board, can through their actions be setting strategic directions of the company. An example in case being Intel, in which the decisions of middle level managers decided the transferal of memory company into a processor company (Burgelman [1994]).

These works bring forth an example of another issue that is relevant for current work. Namely, if we spoke of the three-dimensional MNC, there *another set of dimensions* also could be distinguished. Employees can be considered to be on different *authority* positions(when we use any such formal structure), they can have an *action* position(according to different things that they can really do), and an *information* position(according to the information they possess).[4]

The case in Burgelman [1994] is an example of how, as Hedlund & Rolander [1990] state, the positions on these three dimensions may be not-coinciding. In Intel employees could use their position in action dimension and surpass the authority dimension position of superiors for taking a strategic decisions. And, coming back to MNC theory development discussion, we can say that the works of Burgelman again imply no superior, overarching hierarchy. While speaking of MNCs generally we hereby call them heterarchical, as this body of work examined the structural issues in greatest detail.

---

[3] That is they have derived the internal network position from, amongst other things, the position the subsidiary has in its host country local network.

[4] If Hedlund & Rolander [1990] call these three dimensions the base of a MNC for which the other dimension set is a proxy, we continue to operate with MNCs as three-dimensional(with product, geographical and functional properties) and just speak of authority, action and information positions of different actors.

Analyzing a work that argues for hierarchy (Casson [1994→]), it seems that the points raised do not contradict with heterarchy works – for example Hedlund [1993→] argues that the subordination relations vary and may get switched over time, while Casson shows that these relations naturally arise and exist. Accordingly one of the arguments in heterarchy works is that hierarchy is only a special case of heterarchy. Also one conclusion of the work of Casson is that hierarchy might be best fit for the jobs that with the development of technology are going to be performed by computers – that is it will best function if taken out of human hands.[5] [xi]

Next we are going to consider works that, having their origin at the host country viewpoint, address issues relevant for the whole MNC. An example is a stream of research initiated by the Science Council of Canada on subsidiary management. One of Canadian articles, Poynter & White [1985→], discusses organizational slack that develops in the subsidiaries. Slack is defined as the excess of total human resources after a proper amount has been allocated for the current strategy.

The work shows that subsidiaries have a natural tendency to generate slack and that it can be used for undertaking new value-added activities in subsidiary. This early subsidiary work, though, reports the slack being mostly unwanted phenomenon by the MNCs[»], but shows that it cannot be easily dissipated.

Poynter & White [1985→] mention three ways for dissipating subsidiary slack. The first is through international human resource transfers. Which is a very important, but adversative issue for corporations – for many reasons. Firstly expatriates are important transferors of technology between organizational units (Tsang [*1999→]).

---

[5] Heterarchy works suggest that the line of thought might be best suited for Far-Easter countries as it is in better accord with their philosophies. It is interesting to note that hierarchy arguments (Casson[1994→]) in turn state that the (far away?) East might be a better place for the hierarchies to function as the societies are less individualistic and therefore more eager to enter subordination relationships there.

Also, it is often emphasized that corporate culture is an important tool for heterarchical MNCs (eg. Hedlund[1986→])[xii]. But as seems to be implied in Humes [1993→], creating its carriers encompasses breaking different bonds for employees. For better carrying of corporate culture an employee would have to be transferred between different functions of the company (breaking professional allineation); would have to change its position on product dimension; and would have to work internationally. For discussing the last point we can say that of the different dimensions important for a person, two are national and work related dimension in life. In the interest of corporations is to ensure the importance of working. Thereby people would emphasize their corporations mission(as their mission in life, while working) and be parts of corporate culture. But working internationally for very extensive periods would for those two dimensions mean compromising nationality for work.

We indeed see such *cosmopolitian employees* – global specialists and global managers – working through very international careers and in their case we can say that the work dimension is very prevalent. Their existence is also in a way natural, as we speak of accelerating environmental change, life-long learning and relearning, and lessening importance of the nation state as compared to the corporation (cd. Buckley & Casson [1998→]).

Still it is a tendency for most employees to finally settle for their fatherland jobs (Poynter & White [1985→]) and identify primarily with their home country (Hedlund [1986→]). It seems that this tendency can be said to be natural, because lessening our national identity this way would be pushing ourselves too far, perhaps not only too fast*. Subsequently in this thesis majority of employees are considered to be the 'non-transferable assets' of the corporation and thereby one of the very reasons for ownership specific assets' residing in subsidiaries (cd. Birkinshaw & Hood [1998b→]) and for the existence of Subsidiary Specific Advantages (SSAs) (cd. Rugman & Verbeke [1999→]).

As for the dissipation of organizational slack the other two ways discussed in Poynter & White [1985→] aren't also good solutions. Letting people go from their jobs at subsidiaries would equip competitors with capable human resources and not letting those people take initiatives at all would result in loss of motivation.

Consequently the result of Poynter & White [1985→] implies that as long as the MNCs don't let their capable subsidiary managers to take initiatives, disharmonies in subsidiaries arise. This result is very much alike the ones arrived at in the works that consider the whole MNC. In parallel, for instance, the much later work on individualized corporations (Ghoshal & Bartlett [1997→]) speak of outstanding MNCs that (already) are making use of autonomous entrepreneurial frontline managers as vehicles of development.

It therefore seems, that the earlier subsidiary work was simply written in a situation, where the environmental changes had already taken place but the corporation had not sufficiently responded (as it did not allow for subsidiary slack). Consequently, we can see from discussion of this aspect in development of MNC theories, that the subsidiary works can point to the same direction as research on entire MNCs. Both bodies spoke of a need for leaving subsidiaries with space to operate, be it for example for taking initiatives that are discussed in section 2.2. [6] [xiii]

The works of Poynter & White [1985→] manifest another important issue – namely that we must consider MNCs as *multilayered*. For example we can consider subsidiaries and Regional Headquarters (RHQ) that are built upon them. If going back to our synergy creation example in section 1.1→, we therein spoke of US subsidiary, where positive (production, marketing, R&D) outcomes could be created, then Schütte [1998→] writes of production and marketing synergies – in RHQ case instead.

As Birkinshaw & Hood [1998→] note the different layers of management tend to compete with each other for influence. In addition to obviously emerging layerings – WPM and Divisional Headquarters (DHQ), subsidiary and RHQ – we can also say that the Subsidiary General Manager (SGM) forms a layer between the rest of subsidiary and its superordinates. In Poynter and White [1985→] such position placed SGM in crossfire between his subordinates and the HQ and he was therefore an integral part in efforts to resolve the problem of subsidiary slack.

---

[6] It is hereby argued against (what is done for example in Forsgren and Pedersen[1998→]) drawing a harsh line between the works that have subsidiary as their point of view and those that consider the whole MNC. While it is true, that former can be originated by host country academics they often arrive at conclusions that have much to say about the whole corporation. Next to a discussion on organizational slack we may consider the works of Birkinshaw [eg. 1997→, 1998→(with Fry)] on the types of initiatives that subsidiaries take. These works find that through initiatives subsidiaries help to address the three strategic imperatives that MNCs face: imperatives for global integration, local responsiveness, and worldwide learning (cd. Bartlett and Ghoshal[1992→]). Moreover, an example of intertwined nature of the works is the contribution that Roth & Morrison [1992→] make to terminology about heterarchical MNCs. They state that through creation of mandates (eg. World Product Mandate) for subsidiaries, MNC achieves a state of decentralized-centralization.

From the discussion above we are in conclusion going discuss some important aspects for the SGM's position. We are going to begin with a point touched in the middle of our discussion – a point about evolution of MNC theories. The view of MNC as a multicentered entity emphasizes the fact that subsidiaries can do much in influencing their destiny.

To bring one example from subsidiary literature, it is said there to be a matter of choice, whether we speak of subsidiaries being assigned certain roles, or them (choosing,) acquiring strategies for operating (Birkinshaw & Morisson [1995→]). Although acknowledging that the home country operations tend to play a huge part in MNCs' operations (Ruigrok & van Tulder [1995→]) and that there is an ample amount of subsidiaries operating only as implementers of head-office instructions (Birkinshaw & Hood [1998b→]), we still can see changes taking place in that direction (Birkinshaw [2000→]) pointed by the best practice examples.[7]

For the SGM, it is important to consider that he has to deal with all the diversity on the three-dimensional, partially overlapping structural map – with relations originating from subsidiary or coming from outside. Including, for instance, the WPMs and CoEs that exist.

We have to comprehend that very often SGM is the most cosmopolitan employee of the subsidiary. He is therefore an important carrier of organization's culture and, he can also be a person on an expatriate assignment. It is therefore also interesting to view his authority, action and information position. Having possibly many relationships outside the subsidiary might enhance his information and also action position.

---

[7] As we talked about the fact that there is no superior, overarching hierarchy, this can perhaps be taken as one justification for this study to investigating only the SGM layer on geographical dimension.

But also within the subsidiary his situation might be interesting. As we are investigating the most contributive involvement that SGM can have, the different options as stated by Humes [1993→] are ranging from executive head and strong operational coordinative head to representative head. The interesting issue hereby is whether in case the SGM is the former head of an independent country organization, with less authority now, his relationships (information and action position) still sufficiently contribute for problems investigated in this work.

In discussing conflicts of interest we will come back to discussion about SGM's position forming a layer of management between subsidiary and the rest of the company. In Poynter & White [1985→] he was in a position to pass on information about preferable change of affairs and help solve problems MNC faced.

As HQ was setting the state of affairs, but subsidiary worked for a situation that was to be beneficial for the whole corporation SGM could draw upon his subordinates' capabilities and utilize his contacts outwards the subsidiary. We can imagine that SGM may use his intermediary position to influence other conflict situations that arise.

## 2. INVESTIGATION OF ISSUES OF INTERNAL COMPETITION AND KNOWLEDGE SHARING

### 2.1. A classification of subsidiaries

There is a quite substantial body of articles identifying strategies that subsidiaries have within MNCs. These efforts have had different starting points, for example, D'Cruz [1986→] investigating decision-making autonomy and market scope; Bartlett & Ghoshal [1986→] working with local environment importance and subsidiary's unique capabilities; Gupta and Govindarajan [*1991→] investigating knowledge flows to and from subsidiary in the network of MNC.

This heterogeneity has provided grounds for studies that systemize previous efforts, as is done in Birkinshaw & Morrison [1995→]. We can start discussing subsidiary strategies with the first of their three types – the *local implementer*.

Many subsidiaries have positions in MNCs that are of only marginal importance. If a subsidiary is a 'marketing satellite' for its corporation, it has minimal employment and, as possibilities for growth generally depend on existing resources (Birkinshaw [1997→]), only a small chance for development. Subsidiary might also be a ('miniature replica' or) 'branch plant', that is, a unit that undertakes the entire relevant range of operations in the host country, but does it strictly under specifications from parent company (Birkinshaw [1995b→]).

'Branch plants' do, however, have a tendency to "implement locally", thus deviate from mother-company's instructions (D'Cruz [1986→]), and it is thereafter a matter of controversy, whether they have exhibited the Not-Invented-Here syndrome (Bartlett & Ghoshal [1992→]) and lack understanding of the extent of globalization in their industry (Porter [1986→], Jeannet [2000→]), or have been truly innovational (as for example in Birkinshaw [1995b→]).

But for example in North America 'branch plant' operations were pre-eminent in 1950s and 1960s. Such operations were established in order to provide local content and placate host governments, or obtain relief from import tariffs. (Birkinshaw [1996→]) Globalization, and in particular the dismantling of tariff barriers brought on a process, whereby MNCs reorganized on regional or global basis. This process meant a narrower product and/or functional scope for subsidiaries, but greater volumes. (Krajewski & Blank & Yu [1994→])

The, as we can call it, rationalization-integration of MNCs became a proving-ground for ideas of our local implementers, where only worthy innovations prevailed. The rationalization process has proceeded more painfully in Europe (Birkinshaw & Fry [1998→]) across its many countries. Due to a shorter time span, the operations in Eastern Europe, especially the many acquired companies, are even further behind.

As the research question indicates our objective is to study only 'substantial' subsidiaries, and also those, which are operating in global industries already in an integrated manner. Thus we no longer consider the local implementer type of subsidiaries and hereafter discuss other types.

The two types that are important for us are closely akin in the typologies of Birkinshaw & Morisson [1995→] and Roth & Morrison [1992→]. We are hereafter primarily based on the latter paper, as it deals with the dimension most relevant for us and concentrates exclusively on those types.

The work of Roth & Morrison [1992→] contributes to the strategy discussion by concentrating on the extent of autonomy that subsidiaries have. Subsidiaries can be *Rationalized-Integrated* (RI) into MNC, possessing little operational or strategic autonomy. The other option is for the subsidiary to be mandated.

In addition to *World Product Mandates* discussed, we can see from literature that other groupings, than just a single product or product line, can be considered. Humes [1993→] speaks of moving divisional headquarters away from home country and Rugman & Douglas [1986→] consider mandating a strategic business unit (SBU) to a subsidiary.

As mandates have autonomy in their decisions (from operational to strategic, for example in the case of SBU), they can be viewed as another side of the coin of Rationalizing-Integrating. They are the controlling part in the relationship.[8]

It is worth emphasizing that, as was implied in the discussion of MNCs' divergent structural map, one has to allow for a sub-subsidiary level of analysis. As, for example, different parts of subsidiary have working relationships along different divisional lines, the autonomy level can there also vary (see D'Cruz [1986→], Birkinshaw[1997→]). More recently, it is also evident from Holm & Pedersen [2000→], who note that it has become increasingly important to considered reciprocal interdependencies that subsidiaries have. Thereby one controls some activities, but is dependent in case of others (cf. for example the case discussed in Forsgren & Mathisen & Pedersen [2000→]).

For the purpose of clarification we are going to differentiate the case of headquarters from the above. It can be distinguished by the extent of control possessed over external activities – for example if a product mandate often only controls sales operations concerned, then headquarters of a big division has a range of value-added activities in multiple countries.

But also the ratio of being controlled/being controller differs as headquarters is dominantly the controlling party and not as much interdependent. As a conclusion we can note that by discussing some works, and a systematization effort, on subsidiary strategies we have picked out the most relevant dimension – autonomy – for our further work. We also discussed some basic selection criteria for picking subsidiaries to study as situation of MNCs is changing.

---

[8] It is not predetermined, however, that the mandated subsidiary will be better off than the rationalized-integrated one. For example D'Cruz [1986→] speaks of instances, where subsidiaries are assigned WPMs for products that are at the tail ends of their respective life-cycles, and also brings as an example the Big Three automakers' (General Motors, Ford and Chrysler ) subsidiaries in Canada, which have very favorable assignments to fulfill.

## 2.2. Internal competition in MNCs

If day-to-day operations in subsidiaries can be considered as static, then dynamism in subsidiaries' existence occurs, when subsidiary develops its capabilities and thereby enhances its position within the corporation. From an emergent line of research that considers the development of subsidiaries (eg. Luineks [1999→]) we are specifically interested in the process of capability enhancement. As is evident form ensuing discussion, in the working of MNCs the best way for ensuring enhancement of capabilities is for subsidiary to take entrepreneurial action.

As entrepreneurship in MNCs has been in the focus of interest (eg. Ghoshal & Bartlett [1997→], Birkinshaw [2000→]), both of its forms have received attention. If MNCs can use internal corporate venturing (Burgelman [1983→]), by creating divisions with a specific mission for innovation; then we are more interested in the phenomenon of intrapreneurship (Ghoshal & Bartlett [1997→]), whereby all members of an organization are expected act on emerging opportunities.

Birkinshaw [1997→] has studied entrepreneurial action that subsidiaries are engaging in, namely initiatives that are taken. It is interesting to note that initiatives are primarily distinguished by the level of autonomy that facilitated them. A high level of autonomy (and correspondingly* low parent-subsidiary communication) provides means for external initiatives on the marketplace, whereas low autonomy (and high communication) is characteristic of internal (within the MNC) opportunity spotting and subsequent gain of locating the operation to the subsidiary.[9]

It has been widely documented that subsidiaries of MNCs engage in internal competition (eg.* Krajewski & Blank & Yu [1994→], Galunic & Eisenhardt [*1996→], look relatedly: conflicts of interest between two subsidiaries in 1.1.), and it is easy to see, how initiatives provide means for that. On the one hand internal initiatives exactly comprise out-competing a sister unit for an activity, already performed or upcoming. On the other hand competitive position can also be enhanced by external initiatives, when they extend the market scope beyond host country (in the process also possibly manifesting subsidiary's early recognition of globalization potential in the industry – cd. Hout & Porter & Rudden [1982→]).

---

[9] The level of autonomy should not stay permanently fixed, however. An example can be seen in a case discussed in Birkinshaw [1995→] about a subsidiary unit that produced fan and limit control devices in furnaces. It was at first manufacturing at small volumes. As managers discovered inefficient operations of producing its most complex component (switch) in home country, they redesigned it and the subsidiary gained that activity. Years later it subsequently succeeded in gaining rights for the whole volume of that product – and thus gained autonomy.

Study of such processes in corporations has led to MNCs being defined as operating with internal markets (Birkinshaw [2000→], see also similar points in Buckley & Casson [1998→]) in which corporate headquarters' role is to define the rules of the game and also make certain corporate level strategic decisions. Increased competition is said to be evident from broad use of internal benchmarking and performance league tables, or operations of internal investment agencies.

We can comprehend the competitive atmosphere that arises, as the initiative taker in the case described in previous end-note (fan and limit control device) professed that he was certainly not welcome to the home country unit after successfully challenging their production of switches. In conclusion a final point has to be made about SGM. Namely, its role in achieving subsidiary's competitiveness has been observed to be great, as he has usually been an important driver in initiative processes (Birkinshaw [1995b→]).

## 2.3. Centre-of-Excellence type knowledge sharing in MNCs

Alongside conceptualizations about internal markets that are run in MNCs, we are witnessing a great emphasis that is being put on achieving widespread sharing and trust in organizations. When in the process of developing subsidiary capabilities leading edge solutions arise, it becomes a natural objective to disseminate these throughout the corporation (see for example Andersson & Holmström [2000→]).

In the literature about knowledge transfer and organizational learning great emphasis is put on development of effective information technology solutions and work of mobile managers, who travel 200 plus days a year 'cross-pollinating' ideas and are role models for reaching across organizational boundaries. Organizational culture needed for such processes is characterized as open, based on fairness and shared values.

While surpassing examples can be accounted in service industries – for example consulting company McKinsey– industrial firms such as IKEA are also very impressive. (Ghoshal & Bartlett [1997→], but see also Hedlund & Ridderstrale [1995→]* for projects of international cooperation and )

For analyzing such best practice transferal, we are going to use the concept of Centre-of-Excellence. The subsidiary mentioned previously (in section 1.1.) was 3M Sweden, which had leading edge capabilities in customer focused marketing and key account management. Subsequently being recognized for them and actively helping out other units (Birkinshaw [2000→]), this subsidiary exemplifies a knowledge disseminator in a MNC.

In this work the term CoE is strictly differentiated from WPM. If the latter exemplifies competitiveness of subsidiary, the fact that it has gained itself the right to be the proprietor of certain activities; then CoE is for us a manifestation of the sharing that is also to take place.

It should be noted, that as may be the case with relatively new terms (for example also to some extent true for WPM), it has not yet been refined and a latitude of definitions exist. For the reasons mentioned, our use of the concept is more in line with Moore & Birkinshaw [1998→], than with Forsgren & Mathisen & Pedersen [2000→] or Bartlett & Ghoshal [1986→].[10]

The internal market perspective (Birkinshaw [2000→]) has also discussed the phenomenon of knowledge transfer. In that perspective corporations are said to have internal market of capabilities, one that operates without competition and without fees charged for servicing.[xiv]

Although it is briefly suggested that corporations like HP or Ericsson do a feasible job at sharing at the same time with competing, it still seems to be the point in the internal markets concept worth scrutiny. There is also another work that emphasizes cooperation within MNCs, as opposed to internal market competition. Such is the work of

---

[10] It should therefore be clarified, concerning Figure 1, that mandates can be given exercising control also on functional dimension (R&D mandates for instance).

Eisenhardt & Galunic [2000] on coevolving systems. There the focus is more on sharing and internal competition receives less attention.

Thus at large – the issue of how units (subsidiaries) simultaneously compete and share knowledge is worth investigating. And in conclusion about knowledge sharing it should be emphasized, that SGM's role here is important (similarly to mobile managers). He can utilize his contacts and participation in various task forces or committees to achieve greater leverage of capabilities in MNCs.

The overview of literature has indicated that subsidiaries, having different relationships on the divergent structural map of the MNC, are required to fulfill contradictory roles of being internally competitive and sharing knowledge at the same time. The justified role of the Subsidiary General Manager can be investigated in the light of this balancing effort, as there is previous evidence of this layer of management being active in resolving conflicting situations in MNCs.

While referring symbolically back to Figure 1 we can say that our research problem was to investigate geographical dimension's (that is, Subsidiary General Manager's post on it) role in balancing internal competition, as manifested by a succedent World Product Mandate gainer (drawn on product dimension), with knowledge sharing as displayed by a Centre-of-Excellence (on functional dimension). In investigating this issue, we can say that SGM has an important part both, in internal competition, as well as achieving a state of sharing knowledge.

Our expectation is that in case he has such strong authority position in subsidiary, he can also help to find the balance between the two. Therefore we can put forth a:

*Hypothesis: A strong authority position of SGM has a positive effect on finding a balance between internal competition and knowledge sharing in subsidiary.*

We can expect the situation in various parts of subsidiary (or between) subsidiaries to differ according to the level of autonomy from operations in other countries, as there are direct consequences for position in internal competing. If SGM has had expatriate experience, his action and information positions in the MNC may be enhanced, carrying over to the present issue. Also SGM's information and action position should be considered in the light of subsidiary history – that is, if he formerly headed subsidiary as an independent country organization his previously developed relationships may still be of use.

As the way this process finally maps out is still not clear, there is a need for an in-depth explorative case study to provide answers to questions such as "how" or "why" (cf. Andersson & Holmström [2000⇒]). A qualitative study is characterized by the close relation between the researcher and the entities of the study object (Holme & Solvang [1991]).

Interviews were chosen because of the flexibility of the method. By gathering information through interviews there is a possibility to follow up if needed and there is also space for opinions and ideas from the respondents. Personal contact between the respondent and the interviewer is also assumed to affect the respondent's interest in answering the questions in a positive way. (Bell [1990])

In the next chapter the results of an empirical study of two Estonian subsidiaries of leading European MNC are presented. The multinationals operate in global industries and are already integrated into the network of their MNCs. In total nine interviews, each lasting for about an hour, were conducted. People chosen were managers of different functions in the companies. Questions used were open-ended leaving space for thoughts and ideas from respondents. One of the companies considered it useful to have a number of general questions sent upfront, the questionnaire used is presented in unchanged form in Appendix 1.

# 3. EMPIRICAL INVESTIGATION OF ESTONIAN SUBSIDIARIES

## 3.1. Overview of subsidiaries

We are next going to give an overview of the companies examined in our case study. Thus we will look at the different characteristics that subsidiaries have and the implications for our study. After that we move on to present the case material collected, analyze that and draw conclusions. The information concerning subsidiaries comes mainly from the MNCs' as well as subsidiaries' websites. Additional sources of information are the annual business reports of the companies. Also some information comes from press releases and newspaper articles. Thus this sub-section consists of secondary data regarding the subsidiaries as next presents the primary data (cd Bell [1993]).

A 15 billion $ sales and 880 million $ profit **MNC Akzo Nobel** operates in 75 countries and employs 67000 people. The company is from Sweden and the Netherlands. Diversified conglomerate is active in following areas: pharmaceuticals, coatings and chemicals. Its Estonian coatings **subsidiary ES Sadolin** has 640 million EEK in sales turning 114 million EEK profit. While MNC's main decorative coatings brands are Sadolin, Crown, Marshall, Maestro, and Pinotex, Estonian subsidiary sells Sadolin, Pinotex, Casco, Synteko and Akzo Nobel brands.

ES Sadolin's production facilities of more than hundred (out of 170) employees are located in Rapla. ES Sadolin has ISO9001 certificate of quality.

Although AkzoNobel's is the biggest market share worldwide, it constitutes to only 8% of the total. As the world market is segmented we are going to look at our national markets separately.

ES Sadolin's home market of paints-laquers (of 7 million liters) has three bigger participants: ES Sadolin (ca. 35% marketshare), AS Vivacolor (ca. 35%) and AS Eskaro (13%). As an example of the subsidiary's achievements on home market, in 2001 it achieved the third place in Äripäev's Construction Materials' Producers Top50. Company feels itself to be most akin with AS Vivacolor (Baltic Color AS), a member of industry that is also a MNC subsidiary.

ES Sadolin mainly sells its products on Estonian (25% of the turnover), Latvian (15%), Lithuanian and Ukrainian (together 20% of the turnover) and Russian market, the last accounting for 40% of the turnover. The subsidiary is not solely concentrating on the home market as Ukrainian and Russian markets have highest growth potential. This makes it all but natural that subsidiary has a strong tie for first in home market, but has not achieved a dominant position. In Russia it has positioned itself in the middle niche of the market, the company is active in its marketing efforts advertising Sadolin in national television channel. If in Russian market over the last two years focus was on selling Pinotex coatings then now attention has shifted to Sadolin brand.

The subsidiary can thus be considered MNC's RHQ for greater part of the former USSR. In markets allocated to ES Sadolin its marketing responsibilities cover both self manufactured products and products of sister subsidiaries. Russian market was earlier supplied by Estonian and Swedish products, then after the opening of Russian factory the market is being supplied locally (see below on the factory).

In 2002 the subsidiary also started exporting decorative coating Pinotex to Northern Europe – Denmark, Sweden and Norway. Additional export markets accounted for 14% of the turnover and 150 new products were introduced in the process.

Subsidiary is already integrated into operations of the MNC, as for instance alkyd resins produced in Estonian factory are supplied as a production input to subsidiaries in Finland, Cyprus, Oman and Sudan.

As implied by extending geographical coverage the subsidiary has expanded its production over time. In the beginning of 2002 the MNC relocated part of its Danish operations to Estonia. As a result subsidiary's turnover rose greatly – 25% when compared with the previous year – making that the second straight most successful year ever.

Subsidiary has also expanded in other functions besides production. In the middle of 2002 ES Sadolin received 10 million EEK in investments from mothercompany with an objective to develop its R&D laboratory in Rapla.

The R&D structure of Akzo Nobel consist of a few central laboratories which carry out the fundamental research (in Netherlands, Great Britain and the USA) and regional R&D units which apply fundamental research in order to create products. The goal of the subsidiary is to develop into a development center for our region (although next to Eastern Europe and Russia South-Eastern Asia and North Africa are also mentioned as recipients of services) as not all western concepts are directly applicable here due to cost considerations.

The core competency of the subsidiary is woodcare products with another goal in developing alkyds. There, the consistency of organic solvents will soon be regulated in EU, which calls for development of new types of solvents. R&D department that employs 27 scientists from TTU recently received an addition of two scientists with high academic degree.

Part of subsidiary's expansion has also been geographical expansion of production. In June 2002 ES Sadolin started operating a factory in Russia, near Moscow, which will produce decorative coatings (water-based paints). The annual volume of the factory is 10 million liters. The factory is part of ES Sadolin's business-unit structure; the newly appointed CEO will spend more than half of the year in Russia. The factory is expected to break even in 2005/2006.

Russian facilities cover an area of 7300 m² and include laboratory, warehousing facilities, administrative offices and manufacturing; the staff employed is expected to increase from 50 to 80 over the next three years.

From Russian operations we can also see an example of R&D performed in the subsidiary – as there is a brand that was developed specifically for production for Russian market.

Besides expansion of its sphere of influence within the MNC, the subsidiary also has cooperative relationships with other sister units. As an example we can consider another instance of R&D – cooperation with Italian subsidiary. The result of that project was development of a line of coatings used in renovating facades of historical buildings.

According to Reile ES Sadolin has benefited a lot from information of the MNC – in areas of marketing as in technology – implying cooperative movement of knowledge. Although there is a need to follow MNC's policies, Reile says that ES Sadolin is free in its decisions, thus opening a possibility for competitive initiatives.

French **MNC Saint-Gobain** is amongst the 100 biggest companies in the world recording 30 billion EUR in sales and one billion in EUR in profits. Its subsidiaries reside in 46 countries employing 170 500 people. Saint-Gobain is active in the following areas: glass, housing and high performance materials. It is the worldwide and European leader in all of its business areas. The MNC's peculiarity in the study is in the fact that it has four loosely associated subsidiaries in Estonia. The MNC has invested 470 million EEK creating over 300 jobs.

We are mainly going to be concerned with the oldest **subsidiary** – **Saint-Gobain Sekurit Eesti AS** (SGSEST)– that was created in 1989 in Elva. Its sales are 140 million EEK, profit 30 million EEK and it employs 160 people. The subsidiary belongs to Saint-Gobain Sekurit that is active in automotive glazing.

Sekurit produces mainly windscreens and side windows for the spare parts market of cars. Two thirds of the sales come from windscreens. On the world market the biggest producer of automotive glazing is Pilkington of UK (Represented by Klaasiteeninduse AS in Estonia). In Estonia Saint-Gobain is more prevalent as it produces more for European cars.

There are 300 types of windscreens and close to thousand types of side windows in production at once (for basically all more common cars). Producing such a big number different parts assumes very good logistics within the factory and also control over the technological process.

While describing the subsidiary's role in the MNC and its strategy Kasak has said that Sekurit is a flexible subsidiary that is able to effectively produce small runs of production. Subsidiary produces mainly for European market, but some of its products also reach America, Asia and Africa through intermediary warehouses of the MNC.

The biggest clients are Mercedes and Renault (but also, for instance, some models of Porsche), which send products to their chains selling spare parts. The MNC's subsidiaries in the respective area are active in 24 countries around the world with central warehouse residing in Belgium. Sekurit exports 96 percent of its products (95% of windscreens and 99% of side windows) – to 15 countries.

In 2002 the MNC invested 25 million in order to renew the production of windscreens. As there is a trend in the industry to provide higher value-added products (including mirror clamps and rain sensors etc.) in the beginning of this year the subsidiary started making extruded glass, which attaches to the car with a polymer that acts also as a gasket. The area of production facilities allocated for producing windscreens was doubled in 2002 and correspondingly the number of laminated windscreens produced doubled.

In total the project of expanding the production of windscreens has cost 70 million in investments. The cumulative investments of mothercompany have been over 200 millions.

Previous to that an expansion of the subsidiary took place in the second half of 2000 as it subsidiary started operating a new 70 million factory producing side windows (including annealed side windows). As the technology used for producing annealed side windows differs totally from that used for producing laminated windscreens the expansion process was complex. When this factory opened subsidiary also started cutting glass, as before production inputs were already pre-cut to meet specific measurements.

At the time of this production expansion Kasak underlined it being a result of the great confidence that the MNC has in Estonian subsidiary also mentioning the cost effectiveness of producing in Estonia. When the factory opened operations Kasak also spoke of plans of starting to produce rear windows. He stated subsidiary's objective to be doubling 1999 94 million in sales in the next couple of years.

Also as an expansion of production the subsidiary has for two years produced glazing for Volvo's road construction vehicles. This production input does not go to spare parts chains, but directly to clients' production process. The subsidiary started production of laminated windscreens started in 1991.

Subsidiary's production facilities cover an area of 12 000 m², side windows facility allocating 5000 m². From 1998 the products carry mothercompany's brandname Sekurit. Production of windscreens and annealed side windows is certified according to ISO 9001 – 2000 standards. Annealed side windows and laminated windscreens comply with European regulation R43 and safety standard ANSI of the USA.

Isover Eesti is a selling outlet of insulation materials with 137 million in sales and 2.4 million in profits. There are 20 employees. It also sells suspended ceilings and is active in Tallinn and Tartu. Products are brought from Finland and Sweden, but also from the Great Britain, Poland and Czech Republic. The MNC leadership in its respective markets has carried over to Estonia.

Autover Autoklaas is active in installing automotive glazing (mainly in Tallinn and Tartu) with sales of 25 million, profits 1.9 million and 18 employees. It resells Sekurit's products and also implements projects in cooperation with insurance companies Seesam and ERGO. Company also delivers Sekurit's products to other car dealerships and other Autover companies in Baltics, exports accounting for 15% of the sales. In 2001 company invested 7.5 million to a new dealership.

Baltiklaas produces insulated glass and has sales worth 105 million and 70 employees. The markets of subsidiary are Estonia, Latvia and Skandinavia. The total investments of the MNC into Baltiklaas is 40 million. From the beginning of the year company also produces annealed glass, this product is intended for intra MNC use and moves to other subsidiaries. In 2000, when sales of the subsidiary were around 50 million it produced roughly 100 000 m² of insulated glass.

## 3.2. Results of case study interviews

Presenting the results of the interviews we are at first going to look at the individual steps of development of **ES Sadolin**. As can be seen in table 1 subsidiary began its operations in 1984 and we are going to start with that situation to show the contrast with the state of operations as of now.

TABLE 1

**Steps of ES Sadolin's development.**

| Time | Facility ench. | Output | Sales |
|---|---|---|---|
| 1984 | Paint plant I | 1000T | |
| 1989 | Alkyd plant I | | |
| | Laboratory I | | |
| 1992 | | | 45 million EEK |
| 1994 | Paint plant II | 3984T | |
| 1996 | Alkyd plant II | | |
| 1997 | Laboratory II | 13013T | 530 million EEK |
| 1998 | Warehouse | | |
| 1999 | | 9317T | 332 million EEK |
| 2001 | Service house | 12305T | 517 million EEK |
| 2002 | ICD | 20000T | 600 million EEK |
| 2003 | Tank farm | | |

Source: Intranet of ES Sadolin.

In 1984 only production took place locally. As in the Soviet system there were heavy constraints on getting convertible currency it was impossible to buy production inputs from Akzo Nobel's other subsidiaries or from any other Western company and sell end products for Soviet rubles. Thus the Finnish sister subsidiary was sent samples of a range of possibly suitable production inputs and assigned R&D responsibilities with a goal of developing ES Sadolin's products.

In 1989 subsidiary started both alkyd resins production and R&D activities. As alkyd resins were previously shown as an example of integration into MNC's internal network, this point can now be further discussed.

Alkyd resins are an intermediary production input for solvent based coatings. In the internal network that the MNC operates, subsidiaries have an ample amount of freedom to choose their own partner subsidiaries. Right now the subsidiary has a threefold distinction for its alkyd products. Alkyd resin can either be used for the production of coatings within the subsidiary (primary market), sold as an intermediate input to sister subsidiaries (secondary market) or sold through open market transactions (on tertiary market).

Next to alkyd resins the subsidiary produces wood care products and paints. These are primarily variable cost driven as adding workforce and increasing productivity enables greater volumes. Production of alkyds, on the other hand, relies on chemical processes taking fixed amount of time in a reactor. The production is thus fixed cost driven.

In 1996 the subsidiary expanded its alkyd production by adding the second reactor. By now volumes have reached 6 million liters, next to the amount of 12 million liters wood care products and paints produced. Due to the fixed costs driving the production the subsidiary begins with satisfying its primary market's needs and uses secondary and tertiary markets only when excess capacities exist. On average the subsidiary exports 2 million liters of alkyds.

When an MNC enters emerging markets, such as has been the case of ES Sadolin, there at once open up necessities for cooperative development (cf. eg. Kronzell & Übi & Danell & Kivistik [1999]). Throughout the steps of its development ES Sadolin has also benefited much from the know-how of the mothercompany. First and foremost, it produces coatings that are sold to the end user under Akzo Nobel trademarks. Thus it has 'imported' technology – using the same equipment just as well as receipts for making coatings.

An example of know-how being received can be brought from marketing department's development, though finance has also received help. In marketing, producing the first coatings besides USSR's oil based coatings gave the company a good head start. First marketing works were created basically unmodified from Swedish and Danish campaigns. Marketing developed as corporations from Sweden and the Netherlands merged. Marketing was reporting to Sweden and also received an expatriate from there in the middle of 90s. This employee had very big (Swedish) influence on the way marketing was carried out, and she created basic principles some of which are still in place. Thus in 1994 ES Sadolin was amongst the first (with Saku and Coca-Cola) to start brand building. The corporate training programs attended throughout the development basically are of generally educational nature. In around 1998 marketing concepts were reviewed by Estonians, the focus had also started to shift to Russian market. In 1998 the MNC reorganized and marketing was now guided by international marketing team from the very competitive market Great Brittain. International marketing team gives the subsidiary latitude of freedom – as only things like trademarks and logos are under corporate control – and also provides new ideas. Still it is customary to share some of the responsibility by coordinating bigger campaigns.

There is very much know-how available within the MNC marketing wise. Thus, for instance, one successful marketing campaign was created after employees had gone to Danish subsidiary and reviewed the marketing archives there, finding an exactly applicable body of work.

Another area, where ties with the mothercompany are of benefit is the possibility to use different corporate level software packages. Examples of those are R&D software and software that handles materials safety card, automatically adjusting for regulatory acts updates made in different countries.

But the above discussed production of alkyds for the internal markets also brings us to the competition issues within the MNC. As generally effectiveness and quickness of fulfilling orders are key on the MNC's internal markets, ES Sadolin considers maintaining its price-quality ratio a 'backbone' to being competitive in its secondary alkyds market.

The possible competition within the multinational always has two sides to it. On the one hand there is always a possibility to become competitive cost-wise and thus earn a right to certain production operations. On the other hand there is always corporate politics involved, ensuring that operations in different countries do not completely cease to exist, be it already for the sunk costs involved. Thus even as the Estonian operation can manufacture at roughly a third of costs incurred in Sweden, there is still only a gradual relocation of operations. While looking at the steps of development that ES Sadolin has gone through, we see these almost always being results of local initiative. The process of expanding its markets was taken although the corporation had considered it too risky.

A good example of development of operations can once again be brought from alkyd resins. At first ES Sadolin learned producing it from the Finnish subsidiary – as much as necessary to fulfill its own needs. Now as the export has started, Finland is the major importer of alkyds. Estonian operations have proved to be able to provide the necessary quality while remaining at a lower cost level becoming a supplier to the subsidiary once giving the know-how.

In neighboring subsidiaries there are many operations producing alkyds. The production takes place for instance in Sweden and Poland. The Polish alkyd production is an example of a subsidiary being less efficient, as the factory was set up during planned economy period. Production of alkyds will likely cease there.

Another set of operations that is right now taking place in ES Sadolin, is the production that was relocated from the Danish subsidiary. This brings us an example, where different cost factors, resulting in decreasing sales numbers, in the end clearly outweighed the corporate level politics of keeping the investment alive. The operations that by now have been shut down were located in the center of Copenhagen's city. Over time location became more and more expensive. There were also raising environmental standards to be met, also contributing to costs.

Akzo Nobel has an interesting tool for dealing with the competitive performance of the subsidiaries. From the one hand it can be directly measured – for example cost wise. As was mentioned earlier, in the internal network of the company markets are created. But the markets do not only concern subsidiaries being in partnership with each other. For example marketing subunits of the corporation are also given a freedom to decide, from which subsidiary they source the products sold.

A competitive measure used within the multinational is the annual benchmarking. For example top level managers in ES Sadolin are responsible for carrying out benchmarking of logistics and production. Throughout the corporation measures such as productivity, cost price etc. are created. In Estonian subsidiary benchmarks declined after relocation of Danish operations as ES Sadolin started producing some small batches. In the end, when the full effect of increasing turnover was realized benchmarks returned to their good levels.

The peculiarity of Akzo Nobel's benchmarking system is in the fact that it also facilitates the other side of our research problem – knowledge sharing as subsidiaries are parts of the same company. In the multinational there are forums held. Once a subsidiary has proved its superiority in internal benchmarking, forums aim to highlight the 'simple but ingenious' solutions that it has come up with.

While speaking of formations similar to Center Of Excellences, we may say that during the forums 'temporary COE's are created with a goal of disseminating the created 'best performance' examples. Employees from the contributing subsidiary are given expatriate assignments. Thus within the scheme of internal benchmarking the need to find a balance between internal competition and knowledge sharing is addressed. One example of such best performance dispersal comes from the Turkish subsidiary, and on the level of the multinational such examples of cooperation are numerous.

There are also other examples of cooperative sharing taking place within the multinational. In Akzo Nobel there are horizontal connections formed (directly between the lower level employees) during the joint training programs abroad, such as process safety trainings ES Sadolin's employees recently took part in. ES Sadolin also takes pride in the fact that in a recently organized worldwide training program concerning marketing on emerging markets subsidiary's employee was asked to contribute his expertise as a best practice case. ES Sadolin's marketing manager has been asked to contribute with his lectures on Business Principles, including Business Ethics, etc.

The multinational also forms permanent horizontal bodies – committees – to address business needs. Thus, for instance, marketing manager of wood-care products takes part in biannually meeting committee of colleagues, where fresh ideas, changes in packaging standards and new products are discussed.

The top R&D manager of the subsidiary heads a group responsible for alkyds development on the level of Decorative Coatings throughout the world. The group's responsibilities include providing accessibility of information concerning latest developments and acting on local needs in order to provide responsiveness. The group mainly communicates electronically, but also tri-annual meetings are held. Another example of cooperation in R&D is the Decorative Coatings R&D managers' working group, where Estonians also participate.

When needed lateral connections help overcome critical situations. For example as in case when in another subsidiary one part of production operations had come to a halt and Estonian subsidiary produced the necessary output and sent it to its destination within the multinational.

While considering the structure of the multinational we saw different layers on which the internal market operated. As the subsidiaries were able to choose their own partners for intermediate inputs, marketing units were on a lower level free to switch the sources of products they sell. Another facet of Akzo Nobel's structure is the differentiation between the legal structure and the operative structure. An example of this is ES Sadolin's relationship with the sister factory in Moscow.

Legally, that is, ownership wise ES Sadolin and Dekor are entirely separated, both being owned by the mothercompany. But as was said in the introduction Estonian subsidiary is a regional headquarters for the greater part of the former USSR market. Akzo Nobel uses its operational structure, or 'virtual structure', to create the Baltics-Russia-Ukraine sub-business unit (SBU). On such virtual level ES Sadolin's employees manage the region. Thus for example a person from Estonian subsidiary is assigned chief-executive-officer duties of the Moscow facilities and marketing personnel reports directly to ES Sadolin's head of marketing.

As an example of the structure of a legal entity and its relation to virtual structure we can consider the structure of Moscow's subsidiary (which is in its essence same as ES Sadolin's structure) and subordination relations of SBUs marketing function. The legal entity in Russia has following departments: Marketing, Sales, Site, Finances, Human Resources, General Administration. One employee working in the marketing department is at the same time part of the Baltics, Russia & Ukraine sub-business unit. The marketing in SBU level is basically divided into two: product management and advertising. In product management maintenance technicians are from Estonia and so is for instance the manager of product Sadolin – which is produced in Rapla. But the Russian employee (from Dekor's marketing department) manages marketing of the product Maestro, which is produced in Moscow. The other side of SBU marketing – advertising – divides responsibilities on geographical basis, and there the same Russian employee also has general advertising responsibility for her country. Employees from Latvian and Lithuanian legal entities, for example, have responsibilities for their country.

At large the structure of Akzo Nobels Coatings Group is divided into business units product wise. Sub-business units are constituents of thereof (see Appendix 2). ES Sadolin's Baltics, Russia & Ukraine operates alongside such SBUs as South-America (Brazil and Argentina), Asia-Pacific (Indonesia, Vietnam, China), Central-Europe (Czech Republic, Poland, Slovakia) and South-Africa (in a separate business unit because of cultural reasons) in Decorative Coatings International business unit. In contrast to that geographical orientation Decorative Coatings (Western) Europe is segmented on functional basis. Akzo Nobel also uses product dimension separation – for instance as in Resins business unit in Chemicals Group. The business units in the Coatings Group include: Car Refinishers, Industrial Finishers, Industrial Products, Powder Coatings, Marine and Protective Coating, Decorative Coatings Europe, Decorative Coatings International.

As ES Sadolin supplies the region with the MNC's products, production in Estonian and Russian factories does not coincide. The Russian operations add the production of water-based paints to the region. Already during this year two new products especially for Russian markets have been introduced. From one side ES Sadolin considers Russian factory to be colleagues. On the other hand the two plants also have a kind of a potential competition going on between them – with a drive to perform better and enhance their value chain.

The fact that Estonians were given an opportunity to operate the Russian factory is an example of development of the subsidiary. The Russian factory was not profitable at first, also its trademark was not well known. People at ES Sadolin have daily contacts with Russian factory. Right now Russian factory has a very good quality control laboratory and the transfer of receipts from Sweden has been successful. It should also be mentioned that the person building up the Russian factory for four years was an Estonian expatriate.

Estonian operations have also expanded by starting to produce for North European markets. The tempo of growth of the subsidiary has been quick. Mothercompany has invested ca. 150 million EEK over amortization of money here, adding know-how to that. There have been no unsuccessful expansions in ES Sadolin during its existence.

For production to Western markets (that makes up 30% of Rapla factory's volume), receipts of products have been learned by the plants, thus in that area, no R&D in traditional sense has taken place. Also the trademarks of the products sold do not belong to marketing units of ES Sadolin. In this sense one could call production to North Europe's markets a subcontracting to sister subsidiaries. Subsidiary hopes to stay successful in expanding its production on expense of Northern Europe, based on its indicators.

The highest growth markets – Russia and Ukraine also represent a possibilities for the subsidiary. Thus, for example, ES Sadolin has right now land in excess capacity, as growth is anticipated there.

ES Sadolin's marketing department mainly sells products from such subsidiaries as Finland, Sweden, Poland, and Turkey. But smalls sets of decorative paints also come from countries such as the Netherlands. Both levels of internal market – marketing unit and intermediate product sourcing – work as basically open markets. If prices dictate outside partners may be used.

As a general rule partners are found within a 1000-kilometer radius – due to transportation considerations. An example of a partnership formed is the fact that the SBU sells one traditionally strong Polish trademark in Lithuania, with production of which it has nothing to do with. Another example is a complex arrangement, in which the paint is produced in Sweden, transported to Estonia and then re-decanted and packed here.

There are several ways for forming partnerships on internal markets. According to managers of ES Sadolin during the first four to five years of employment one can build up a reliable network of contacts – as top management in the multinational is stable over time. One can use contacts of ones contact person. Also, there is a good possibility to find such information from the MNC's intranet. In addition there are ombudsmen assigned for such purposes in the corporation.

It can be said that R&D units are also operating on internal markets. As the structure of Decortive Coatings International is geographically based there is R&D work that is doubled and there is a necessity to specialize further. But already there are different subsidiaries having strength in different areas. Thus for instance the subsidiaries of Turkey and Brazil are advanced in paints, as ES Sadolin is strong in wood care products, also alkyds.

In R&D market, there is a possibility to develop competencies of subsidiary and subsequently receive corporate level investments. After that other subsidiaries will start buying R&D services from the developer. This is also a tool for achieving specialization of R&D. An example of providing services is the R&D department of Danish subsidiary, which has specialized after the production relocation. Now it provides corporate clients with research concerning microbiology. (R&D services are paid for be becoming a co-contributor in yearly budget of the lab or sometimes also through one time bills. This also depends on weather the lab is a part of producing subsidiary or just standalone.)

There are a number of corporate level Centers Of Excellence – such as for wood care products, toners and laquers. ES Sadolin's wood care services are perhaps not as up-to-the-maximum sophisticated as the respective corporate (Western-European) COE's, but can well be applied and are much less costly. The cost for R&D services is calculated on man-hour basis. Therefore in the general process of research consolidation a goal of becoming wood care Center Of Excellence. As Estonians basically perform the same tasks as the COE it can be said that there is a healthy competition in the MNC. Units strive for better results and faster development speed. China, Indonesia and Vietnam bring us an even deeper needs to economize costs and thus there, for instance, research is also carried out locally.

Generally speaking, information concerning R&D is fully available within the multinational. Thus if one subsidiary finds it to be in the business interests of the MNC to come to its market with some product (from any of the MNC's units), it can also manufacture it. Therefore the R&D manager seeks authorization from the business unit level functional head, or some higher-level functional head if necessary.

Thus, hypothetically speaking, if Hungarian manager finds that a paint has can be sold, raw materials effectively procured and local costs kept down; he arranges for receipts to be transferred to the subsidiary to start production. There will, however, be corporate level restrictions on entering other markets that sister subsidiaries already supply. Also there are some restrictions to what brand names can be used.

When we compare possibilities of development of a subsidiary it can be said that R&D is more specific and complex. Therefore the actual production operations are more of a matter of 'proper effective-and-consistent execution'. This provides for footloosness of production. For this reason one can also see that sub-business units protect their production and try not to let is 'slip away'. Thus here we can see that relationships between subsidiaries can be to a degree competitive.

Estonian R&D department has been a corporate supplier of its know-how to subsidiaries in Argentina, Indonesia, Poland, Hungary and Turkey. The complementing strengths of Turkish and Polish subsidiaries have enabled ES Sadolin to receive know-how. The joint project of developing renovation paint for historical buildings with Italian subsidiary is also an example of Estonian subsidiary receiving know-how. If we look at the total R&D knowledge flows within the MNC the subsidiary has over its existence received much more information than given.

A final note related to competition between the subsidiaries is from the relocation of production from Denmark as ES Sadolin went through a successful internal initiative (cd. Birkinshaw [1997]). R&D department of the subsidiary managed to transfer and initialise production of 20 0 receipts during one year by maintaining quality and adjusting to specifics of local machinery used.

The process ended well for the people that were let go from the Danish operations as other job opportunities became available during the layoffs. Still, if relationships with the Danes had been good, they worsened during the process, as Estonians were basically the ones taking away the Danes' jobs. Although information concerning such issues as receipts of products and relationships with cans' suppliers are basically a property of the employer, people were reluctant to transfer to Estonians. Process went most smoothly on R&D level as Danish lab was kept operating. More problems occurred on ordinary employee level, perhaps also with mid level managers.

For the Danish subsidiary the uneasiness of the situation is understandable. Denmark is the country were Sadolin was created 230 years ago. Just as well the wood care product Pinotex originates from there. The multinational has a trademark there that in its strength is among top five throughout the country – in any industry. And in such background a 'younger sibling' managed to overtake operations.

As we have seen there have been both competitive and co-operational aspects in ES Sadolin's development. The company is quite free in its decisions, operating in, as Akzo Nobel terms it, the decentralized network of the multinational. Over the years ES Sadolin has, through its financial results, shown excellence in the context of the MNC. This can be considered as means for 'buying' operational freedom within the multinational. As far as the numbers are good all one has to do is to follow corporate level policy. We can consider ES Sadolin to be an autonomous subsidiary in our research as it has gained a reputable position within the MNC over the last six years.

In the automotive glazing world market main competitors are **Saint-Gobain Sekurit**, Pilkington, Asahi from Japan and PPG from USA. When new car models are introduced there is a competitive bidding for supplying rights.

As at first the demand for new products is growing, bigger factories of Saint-Gobain Sekurit perform the production. The volumes produced reach millions of sets. When product reaches saturation phase the production of glazing is phased out to smaller factories, so that bigger factories can retain large volumes.

In glazing, products supplied go to Original Equipment Manufacturers (OEM), that is straight to car factories; to official replacement market, that is to spare parts market through official dealership of the car plants; or to aftermarket under glazing company's trademark. It is possible, that quality standards for producing to different markets vary. (Areas, where micro stain is allowed, standards on allowing dust between layers of glass are examples of such differentiation.) The fact that a product is being phased out eventually leads to the car being no longer produced, meaning that the OEM market no longer exists.

Estonian factory is a niche factory in the corporate context. One reason for not supplying production lines directly is in the fact that subsidiary operates far from car factories. As these companies maintain absolute minimum stocks they need new deliveries on hourly basis, which is hard to achieve from Estonia. Especially given the additional need to cross EU border in Poland. Still distance is not the only reason for the coporate level decision, as SGSEST used to supply car factories in Poland and Russia in earlier phases of its development.

There is also a small-scale production going directly to assembly lines, such as in case of Volvo. In Swedish factory a decision to cease production for road construction vehicles was also made in order to maintain big volume production. SGSEST yearly produces only around 11 000 items for Volvo. But as, for instance, producing for one type of Mercedes, output is directed to official aftermarket.

As can be seen from Saint-Gobain's organizational chart depicted in Appendix 3, there are a number of factories operating under Transportation Division. The sales of spare parts market are handled by another unit of Saint-Gobain named Autover. Autover has warehouses in different countries. In earlier phases of development the goal of SGSEST was to penetrate new markets and obtain a market share there. Right now these operations are parts of Autover, thus we can say that SGSEST mainly has one client that operates in intra-corporate market.

Autover's multiple warehouses have direct contacts with Sekurit's different factories. SGSEST sends its products to Belgium, Sweden, Poland and Austria, for example. In ES Sadolin's example we saw that internal markets of the multinational were potentially opened. In SGSEST's case we can also bring examples of openness from cases when different subsidiaries have not matched expected price levels. Autover has then made purchases from independent producers.

To evaluate performance of Sekurit's different factories the MNC, just as Akzo Nobel, uses comparative benchmarking. Next to (monetary) controlling performed in subsidiaries all non-financial characteristics are gathered in a system named TABIA. TABIA shows indicators concerning quality, stock, warehouse, production outturn, raw material yield; and man-hour-, direct-, indirect-, general- and machine specific productivity. Thus everything that can be measured in square meters, items, percentages and ratios is a part of TABIA. The system provides a comparison of operations located in different countries.

TABIA brings us to a measure of control within the multinational. When compared with Skandinavian MNCs (as Akzo Nobel also has Scandinavian roots), which tend to be decentralized, French MNCs usually exercise a more tight central control. Such is also the case of Saint-Gobain. Thus, for instance, financial management of the subsidiary takes place not through biannual but monthly reporting. Also on monthly basis TABIA files are sent to Paris, which consolidates reports of around 40 plants. Upcoming investments are decided way ahead, as they are at first introduced into five-year business plans and go subsequently through all shorter term plans.

Although SGSEST's reporting structure is tighter than was the case for ES Sadolin, there are no major implications for the autonomy of the subsidiary. This subsidiary is also autonomous, although ES Sadolin was especially so – operating its own geographical sub-business unit. When comparing the orientation of initiatives that subsidiaries take we will subsequently see, that if ES Sadolin was oriented more towards external markets (external initiatives), SGSEST is right now internally oriented (taking internal initiatives within the MNC, although orientation towards new markets has also been a characteristic of subsidiary in its earlier development phases, as discussed earlier).

After the beginning of production in 1991 main steps of development of the company have been starting the production of side windows and the expansion of windscreens production. Still the growth of the company has been continuous. It has also taken place by increasing the number of shifts at work, and subsidiary also varies the number of days worked by a single shift. Subsidiary's and its general manager's initiative in achieving the growth of company have been of decisive importance. For example as the production of side windows started, SGSEST won the investment while competing with the subsidiaries from Poland and Portugal. When the production of windscreens was expanded some operations formerly taking place in Sweden were divided between Poland and Estonia. Good financial results have provided the possibility for these decisions that are made on the level of Saint-Gobain Sekurit's management with also Estonian general manager participating.

Estonian subsidiary has following departments: procurement, production, sales, quality and environment; and finance. The subsidiary performs only production and thus development, but no research takes place. As the production at first takes place in central factories, R&D laboratories are in contact with those while introducing production. Estonian subsidiary has contacts with central factories, not laboratories while the products are being phased over.

When a product is being introduced in SGSEST, employees of plants make mutual, couple of days long visits. There is both hardware and software transferred. Depending on which things are possible to take over transfers are made. For example, bending frames and technical data of products may become of use. Software wise the job of SGSEST's production department is to redo the manufacturing of the glass. The parameters of a glass being its measures, convexity, angles, optics, and internal tensions.

The MNC has contributed during the development of the subsidiary. As production was expanded there was an engineer from MNC's headquarters assisting the technical side of the project. There are numerous training programs taking place, for instance in one central institution in France. Topics covered include management, financial management, costing, internal audit, work safety, marketing, personnel management, logistics, bending glass etc.

There are workshops, where participants are asked to present 'best practice' cases from personal experience, and in this process we can already speak of different subsidiaries exchanging know-how cooperatively. Also temporary working groups are formed – for instance in order to introduce financial software in the multinational, such as was the case with Enterprise Resource Planning.

Cooperation also takes place in technology related programs, the main topics discussed concern glass and folio bending. Such work takes place on sites. When, for example, a new furnace was installed, its operators spent weeks abroad. There have been employees of other plants visiting Estonia, such as for instance a three-day visit paid by Czech employees as a new folio bending device had become operational in Estonia. Also an employee from Indian factory came to receive know-how.

In case of employees of Thailand's subsidiary visiting, SGSEST was the receiver of knowledge about bending folio. In a recent visit to Germany, SGSEST's employees received know-how about bending glass. The technology of making extruded glass was learned from the Czechs.

Due to SGSEST's small volumes, its production know-how is of a somewhat specific nature and there is a shortage of application examples. Thus it might be so that some workshops, where employees of Mexican, Brazilian or German subsidiaries set the tone, are just interesting as examples of how huge factories are operated, but not applicable at home.

Although SGSEST is not able to produce rear windows, it doesn't complicate matters much, as its target is the aftermarket. For car companies it would be much more important to purchase all components from one location. Furthermore, it should be noted, that there is by nature a smaller demand for rear windows producers, as the main sources of damage – small randomly flying stones, damages by break-ins, and car accidents – affect rear windows the least.

Within Saint-Gobain Sekurit also different factories produce for the aftermarket. There are different characteristics determining the success. Besides these mentioned in the discussion of benchmarking, importance lies in quickness and ability to meet deadlines, quality, flexibility, and price.

There are some general rules that warehouses have to follow, which are set through the mandatory lists. Mandatory lists set ranges of products that have to be bought from specific factories. Thus, for instance, the Polish subsidiary is in the mandatory list for Volkswagen cars, DaimlerChrysler's Mercedes cars get their glazing from the German subsidiary and Volkswagen's Škoda cars are supplied by the Czech subsidiary. Exceptions of a few thousand items may be created, but bigger violations are not prohibited.

It is noteworthy that despite being situated in the outskirts of its region SGSEST still has competitive edge. This although its production inputs are transported from Germany and France; and its finished products move back to earlier mentioned warehouses in Europe. Low labor and energy costs as well as taxes also contribute for the efficiency of the subsidiary.

When comparing the amount of subsidiary's orders received with other corporate producers, we see it succeeding. Sekurit's subsidiaries located in Finland and Italy also produce for the aftermarket. Also besides producing for Škoda, the Czech subsidiary is active in production for this niche. There is also an overlap with the Polish production operations as well. Additionally a factory producing windscreens and side windows, just as SGSEST does, is located in India. It produces for the local market, but would also be interested in expanding its operations by targeting other – for instance European – markets.

As a consequence of having overlapping operations and same buying partners, there is also competition between subsidiaries. When we considered cooperation that takes place between subsidiaries, many instances occur with subsidiaries that are not niche producers. Such is the case in contacts with German, French and Swedish subsidiary. Still as already described there is also good cooperation with, for instance, Czech subsidiary.

The nature of relationships also depends on the state of the industry, as in buyers market like the one existing right now, it is harder to obtain orders. Thus for every factory, and different types of operation taking place within factories the amount of orders received directly translates into number of shifts at work and days worked in a shift. Which means that these are the jobs of employees of any subsidiary depending on the ability to perform well within the MNC. When a subsidiary introduces a new product, there is a possibility that in small number of cases cost price levels attained are not sufficient for Autover and the product will not be purchased. Such introductions will on the one hand still mean that subsidiary has a wider range of products to offer for its clients, but on the other hand, if the products will practically not be used (which seldom also happens), the cost of development might be considered as 'garbage' cost.

Concerning relationships, there are instances within the MNC where openness (and free movement of information) with other subsidiaries ought to be bigger. People working in the subsidiary feel there to be a competition between sister units, but this of course is already established in the general framework of the MNC.

In conclusion we again see a situation, where the subsidiary has to find a balance between internal competition and knowledge sharing. In the global car industry this goal is harder achieve than it was in coatings business.

We are next going to take the **subsidiary general manager**s' viewpoint concerning our research problem. In ES Sadolin's case the more competitive situations in subsidiary's history have had their roots mainly in 'objective' rather than 'subjective' reasons. Meaning that for instance the production in the center of Copenhagen simply became very expensive due to various additional expenditures rather than decreasing effectiveness of production. Therefore there is less complexity in finding a balance between being competitive and sharing freely. Also the system of benchmarking the subsidiaries' operations and thereafter dispersing the best practice examples helps to maintain the balance, as well as the fact that markets are assigned geographically to sub-business units. Thus Estonians have, for instance, been free to make visits to Scandinavian plants. In case of R&D, though, the historical COE in wood care is to a degree protective, when higher (managerial) level contacts are considered.

In case a need to balance the subsidiary better would arise there are several lateral connections that SGM has. He takes part in a body governing the business unit. Also there are meetings taking place on the level of the whole corporation, including the meeting of country coordinators where one representative from each country attends. Between these two levels also Coatings Group meetings take place. SGM can additionally become a member of different bodies if he wishes.

In ES Sadolin's case the importance of subsidiary general manager's role lies in balancing the sub-business unit instead. For an employee the contradiction can be said to lie in directing ones efforts towards the success of the legal entity as opposed to operational entity.

An example can be brought from a creation of a new company in one of SBUs countries. As a company would start its operations small there would only be around ten employees. The expectations of ES Sadolin's SGM (who is the head of SBU, acting in its interests) would be for the manager of the new company to devote 70% of his energy into representing all different parts of the SBU and 30% into developing the local subsidiary. The manager in question on the other hand might want develop the local company more.

There are cases, when Akzo Nobel's expansion has created a competitive environment within the corporation. In many countries the multinational has developed by mergers and acquisitions. Thus, after the merger of Akzo and Nobel itself two fierce competitors on United Kingdom market found themselves to be within one company.

There are no such cases in Baltics, Russia & Ukraine sub-business unit. But, for instance, after Estonians started operating the Moscow factory, there were next to already existing complementary brands also those founded on same base receipts. Consolidating by stopping production of those meant losing their creations to developers. As there were already established own products and working client relationships sometimes the tendency was to concentrate on those. Also it was seen, that if a product goes out of Russian warehouse, it is its own and helps the factory start turning profit.

For the SGM the goal is to work through these contrarieties. As the issues are discussed, for example it is made clear, that the assessment of Russian subsidiary's results is based on success of the MNC as a whole and sales of all products are rewarded. The fact that an Estonian heads the Moscow factory helps to achieve the balance. Thus for the SGM there can be no competition between the Moscow and Rapla factory as it is the SBU level that matters. It doesn't rule out the possibility that people in the lower positions in Estonian factory are patriots of specifically their operation.

SGM additionally handles relationships with SBUs from other markets. Thus, for instance, the SBU of Turkey has its salespeople in Russia selling complementary products. There was also similar effort in Lithuania by Poles, but SGM decided that as BU&R is the sub-business unit's marketing area, Estonian marketing started selling the products itself. In case SGMs take such decisions, they rearrange the way profit is divided between BUs. In the latter case as Poles had been receiving full profit at market prices, their situation weakened as BU&R SBU started buying products from them at intra corporate prices. Likewise BU&R transfers its profits to Scandinavia while producing for that market.

Moving on to the case of Saint-Gobain Sekurit Eesti's subsidiary general manager, we see that in his subsidiary the competitive issues arising relate more to the day-to-day effectiveness of the operations. There is a price-based competition for supplying the corporate clients.

An important point to make is that the subsidiary general manager's goal can be formulated as a need to unite the level of organization as a whole – where the MNCs always have to declare unity – and the 'human' level, on which employees have to act in the interest of the job opportunity, their career etc. By successfully dealing with these two levels that exist in an organization, the SGM balances internal competition with knowledge sharing.

In SGSEST subsidiary general manager also has lateral connections, besides the line of command he is a part of. He participates in managerial bodies – on the level of Saint-Gobain Sekurit International, as well as Saint-Gobain Sekurit International's Transportation Division. There are also bodies that are formed between Transportation

Division from the one side and Saint-Gobain Sekurit or Autover from the other (see the relationship on organizational chart in Appendix 3). One example of lateral connections that he has is through Transportation Division's committee.

The organizational culture of Saint-Gobain, influenced by the French origins, is soft and encourages discussion. In case inter-subsidiary problems, that SGM has to address, arise he also uses his information and action position working through persuasion and convincing, not authority.

As was also evident in ES Sadolin's case, the success of (lower level) horizontal connections between subsidiaries is formed on case-by-case basis and depends on personal relationships as well as intercultural communication. The mission of SGM as the most international employee of SGSEST is to revive such relationships in case any problems have arisen. He has to gain accessibility to the information needed as well as address concerns of Estonian employees.

In conclusion we may say that as in ES Sadolin's case the SGM was overseeing a regional headquarters he relied more on his authority position (by making changes in the way SBU was managed); in case of SGSEST SGM relied on his information and action position in order to resolve conflict situations – both working to keep a balance between internal competition and knowledge sharing.

# CONCLUSION

In our work we saw that during the last decades the environment that the MNCs operate in has changed – becoming more volatile and less pacedly growing. In this environment the MNCs themselves have become more complex and also flexible. We found that MNCs are essentially three-dimensional, that is, they organize around product, functional and geographical dimensions and exhibit characteristics that having a common origin can be applied along any one dimension.

Therefore we depicted MNCs as having a divergent, partially overlapping structural map. On that map there can, for instance, be functionally oriented Centers-Of-Excellence; product dimension World Product Mandates; and for capturing synergy of a big set of operations country dimension based arrangements.

Analyzing the development of organizational aspects of MNC theories we saw different bodies of work pointing to a similar direction. We followed developments of concepts Heterarchy, Transnational (and the related Individualized Corporation), works of interorganizational theories school (the multicentered MNC), works considering autonomous strategic decisions, and works originating from subsidiary (host) country research. In addition to the structural developments mentioned earlier these works also point to a need to empower the frontline units.

Outlining our research problem the conceptualization of MNCs as operating (competitive) internal markets was shown to rely on entrepreneurial, initiative taking behavior and result in the development of the subsidiary. We classified the subsidiaries – first selecting operations substantial enough and then differentiating them based on the autonomy level, as it has implications for the types of initiatives taken.

While internal markets concept can be said to put more emphasis on competitive behavior in MNCs, conceptualizations such as MNCs as coevolving systems concentrate more on cooperative sharing that has to take place. Symbolically referring we considered COE to represent sharing that takes place, while WPM was an embodiment of competition, and subsidiary general manager, being on the third – country – dimension, balances the two. SGM has an important part in both processes – as he is an important 'cross-pollinator' of ideas as well as initiative process champion.

Next, the most contributive involvement that the SGM has was researched. A case study was conducted, where SGMs role in balancing the subsidiary between internal competition and knowledge sharing was investigated. Nine interviews were made with managers of subsidiaries ES Sadolin (Akzo Nobel) and Saint-Gobain Sekurit Eesti (Saint-Gobain).

The structure of the MNCs studied had indeed complex embodiments. Thus, for instance, there were Centers-Of-Excellence – as a part of Akzo Nobel's R&D structure, and also the goal of ES Sadolin's laboratory was to become one. There were also temporary COE like formations, with a goal to disperse best practice information. Also different forms of mandating were present: subsidiary producing for markets besides its usual ones (AN), a subsidiary winning a competitive bidding for a new production mandate (SG). There were also numerous instances of cooperative bodies formed with employees of different subsidiaries participating.

The structure of the subsidiary was much more complex in AN's case, as the subsidiary was a regional headquarters. AN's subsidiary represented the full value added scope of operations as SG's concentrated on production mainly.

In both cases the internal markets within the multinational were operational and also open. In case of AN R&D services as well as end- and intermediate products could be purchased from other subsidiaries, as the regional headquarters had gained rights to operate its markets. In SG there were cases of purchases from external market instead of corporate suppliers.

Both subsidiaries can be considered autonomous, still AN's was especially so. It had implications as suggested in theory – AN's subsidiary was oriented more towards the external market opportunities than SG's subsidiary. The level of autonomy suggested that local subsidiaries had indeed been empowered.

As subsidiary is not as internally oriented, and also as the factors behind transfer of operations have been of more 'objective' nature the subsidiary general manager has to do less balancing in AK's case. There was also a system of combining competitive benchmarking with best practice transferals in place. Instead, his efforts are directed towards keeping part of the sub-business unit (region) from taking parochial, non-cooperative point of view.

SG's subsidiary general manager has to make a bigger effort in balancing the subsidiary. In price based competition with sister subsidiaries that produce for the same niche, there are situations where cooperation ought to be bigger. As in his counterpart's case, there are lateral connections that can be utilized in order to achieve the better functioning of the MNC as a whole.

In order to balance the subsidiary a goal that subsidiary general manager has to achieve can be formulated as a need to unite the level of organization as a whole – where the MNCs always have to declare unity – and the 'human' level on which employees have to act in the interest of the job opportunity, their career etc.

There is a difference in means that the subsidiary general managers can use. As a regional head, AN's SGM takes the authority position to better maintain balance between sharing and competing. SG's SGM uses persuasion and convincing – works through his action and information position – for reaching the same goal.

This does to a degree rebut the research hypothesis that was put forth before the empirical investigation. Hypothesis proposed, that authority position would be of use while balancing the subsidiary between countering demands of internal competition and knowledge sharing. In our case the authority position has helped AN's subsidiary head – but to maintain the balance in his region, in operations directly responsible to him only. In the case of needing to balance the operations below a SGM with operations in another sub unit – as was the case in SG, the subsidiary general manager used his information and action dimension positions instead.

# APPENDIXES 1-3

Appendix 1. Questionnaire sent beforehand.

Millised on olnud erinevad allüksuse arenguetapid?

Kas läbi erinevate etappide on initsiatiiv laienemiseks tulnud allüksusest või on määravaks saanud emafirma keskne juhtimine? Kust on tulnud investeeringud?
Milline on olnud laienemise läbiviimise protsess, kes on olnud tegevad?
Kas on olnud luhtunud tegutsemise laiendamise plaane?
Kas läbi tegutsemise on olnud konkurentsi teistelt allüksustelt?

Millised on olnud koostööd nõudvad olukorrad läbi organisatsiooni toimimise?

Millist koolitust on ettevõte läbi toimimise emafirmalt saanud, milline teadmiste ülekanne on toimunud?
Millistes multinatsionaali läbivates harvemini või tihedamalt toimivates tööorganistes (kirjanduses: task force, commitee) osalevad allüksuse töötajad koostöös inimestega teistest allüksustest?
Milline on välislähetusse saatmise poliitika multinatsionaalis, kuidas liigub tööjõud?
Kas on olnud koostööd arendustegevuse vallas?

Millisel kohal multinatsionaali hierarhias asub allüksus ja kui iseotsustav on Eesti allüksus?
Millised on multinatsionaali meetodid allüksuse edukuse (hind, kvaliteet) määramisel?
Kas allüksuse võimaluse tellimusi saada määrab nö. multinatsionaali sisemisel turul edukas olemine?
Kas multinatsionaali sees on ka üksusi, kellele on määratud roll teistele oma teadmisi edasi anda?

Kas konkurents teiste allüksustega on vajalik? Kas koostöö tegemine on vajalik, kas seda tehakse piisavalt? Kuidas leitakse tasakaal koos töötamise ja konkureerimise vahel?

Appendix 2. Organizational chart of Akzo Nobel's Decorative Coatings International business unit. Source: Akzo Nobel's intranet

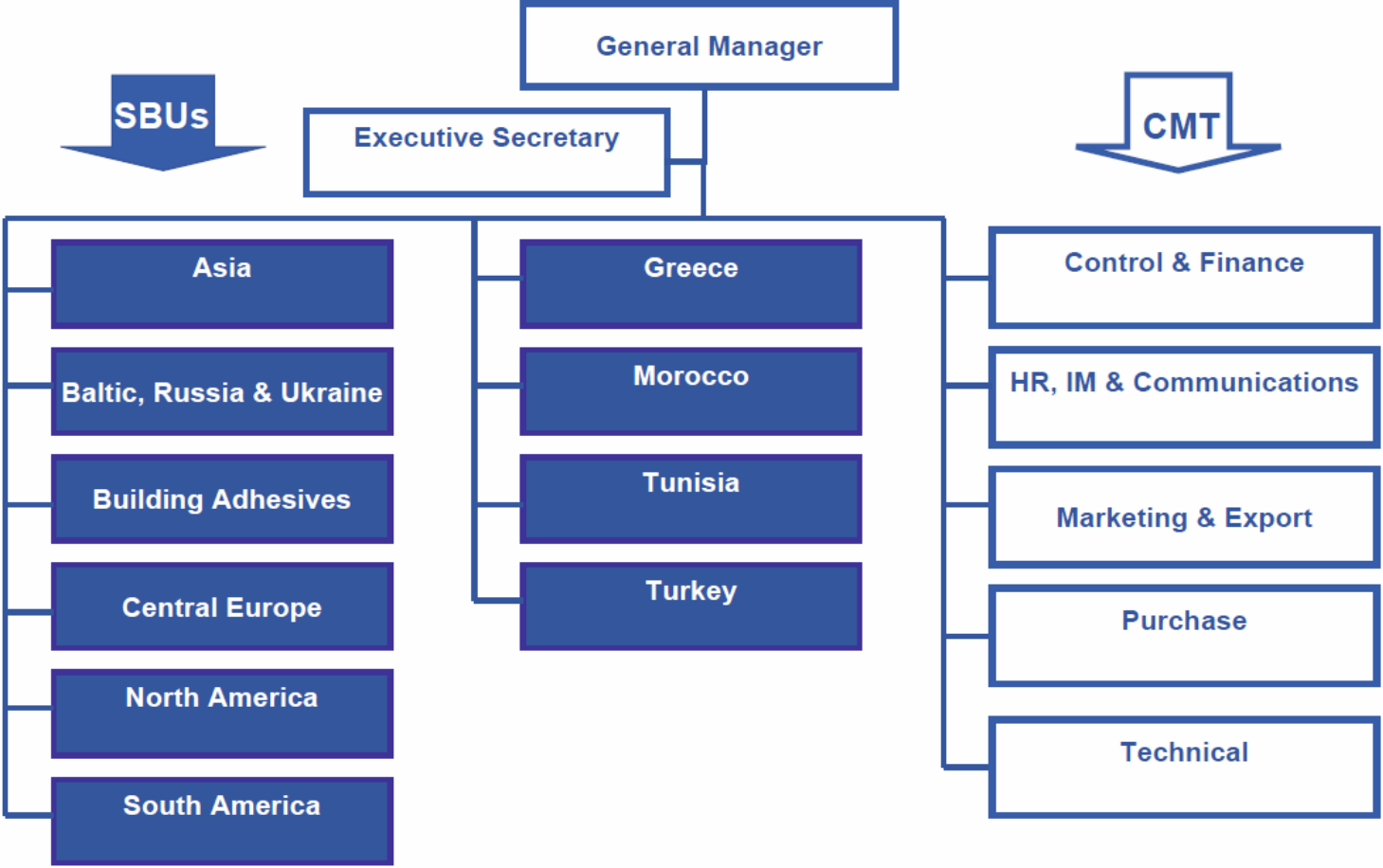

Appendix 3. Organizational chart of Saint-Gobain Sekurit's Flat glass Division.
Source: www.sain-gobain-sekurit-transport.com/instit/orga/index.htm

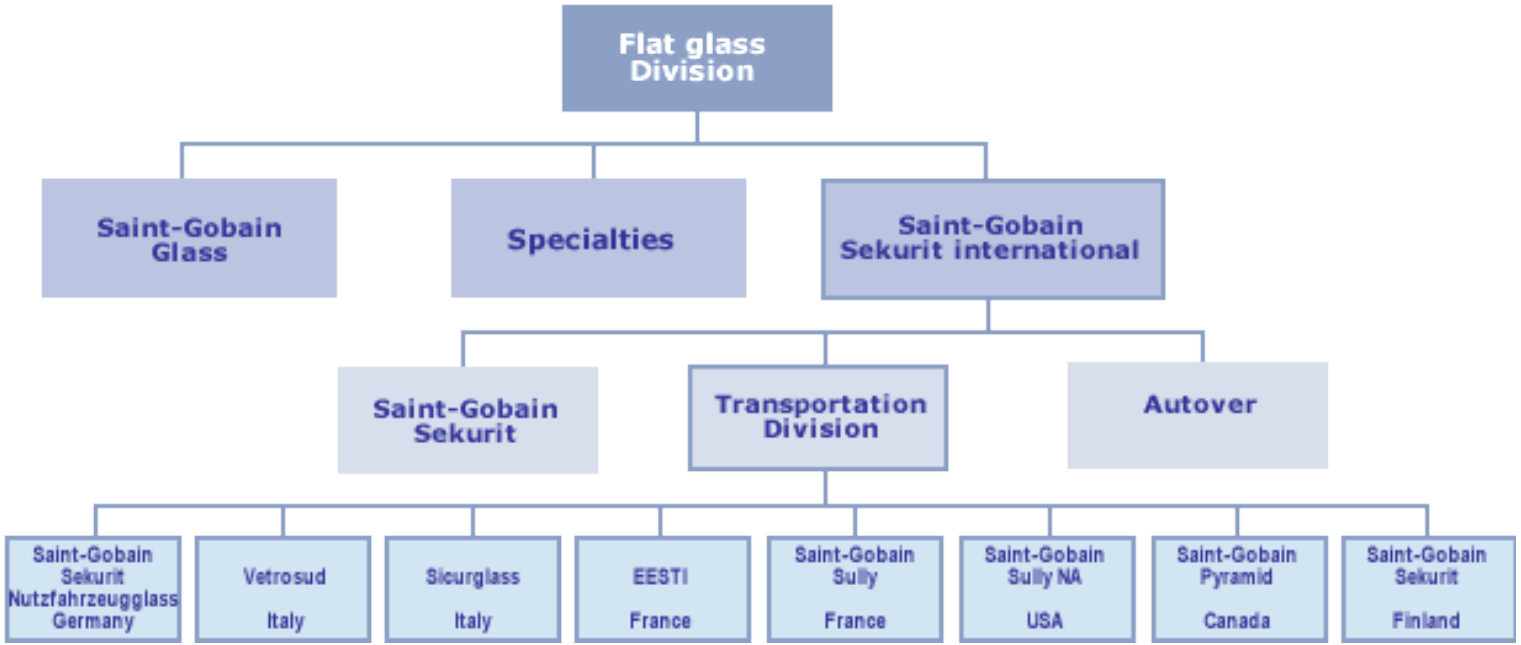

# SISUKOKKUVÕTE

Multinatsionaalsete ettevõtetes toimunud organisatsiooniliste muutuste käigus on viimasel ajal esile kerkinud uurimissuund, mille fookuseks on allüksuste juhtimise probleemid. Sellist vaatenurka põhjendab fakt, et allüksuste potentsiaali rakendamisel on võitjaks ettevõte tervikuna.

Selle üle kui suur peaks olema tippjuhi (SGM) otsustusvõim lähevad arvamused lahku. Antud uurimuse eesmärgiks on vaadelda allüksuse tippjuhi rolli, millega ta ettevõttele kõige rohkem kasulik saab olla. Kasulikkust uuritakse kontekstis, kus tippjuht peab allüksuses saavutama tasakaalu teiste allüksustega konkureerimise ja teadmiste vaba jagamise vahel.

Töös leitakse, et MNCs on oma olemuselt kolmedimensionaalsed, nii et erinevad omadused võivad olla käsitletud ja rakendatud kas toote, funktsionaalse või geograafilise aspekti vaatenurgast. Graafiliselt kujutame MNCd kui divergentse, osaliselt kattuva struktuurikaardiga organisatsiooni. Teadmiste tippkeskus (COE) on näide funktsioonimõõtme prevaleerimisest, samas kui tootmismandaat tootedimensiooni rõhutab; ning juhtumiks kus geograafiline mõõde prevaleerib on suure arvu ühes riigis asuvate operatsioonide sünergia eesmärgil koordineerimine.

Töös vaadeldakse multinatsionaalide teooria arengut, seda kuidas tekkisid praegused struktuurilahendused ning kuidas allüksustele rohkem otsustusõigust antakse. Käsitletakse heterarhiat, transnatsionaali, interorganisatsiooni koolkonda, autonoomseid strateegilisi otsuseid ja asukohariigi poolt algatatud uuringuid.

Uurimisprobleemi raames tuuakse välja nii kirjeldused, mis rõhutavad konkurentsi MNC sees – sisemiste turgude käsitlus – kui ka koostööd ja vaba jagamist – koos arenevate süsteemide käsitlus. Nii konkureerivate initsiatiivide kui teadmiste jagamisel toimuva kontaktide loomise juures on tippjuhil suur roll. Allüksustest valitakse välja suuremad, ning nende vahel tehakse vahet autonoomia järgi, kuna see omab tähtsust MNC sees konkureerimisel.

Kaasusuuringuga tegeldakse kahe allüksusega – ES Sadolin (Akzo Nobel) ja Saint-Gobain Sekurit Eesti (Saint-Gobain) – ja viiakse läbi üheksa intervjuud.

Multinatsionaalide struktuur on kompleksne, esinevad nii COEd kui mandaadid. Töötavad allüksustevahelised koostööorganid. AN puhul on allüksuse struktuur keerulisem, kuna tegu on regionaalse peakorteriga, SG allüksus tegeleb tootmisega.

Mõlemal MNCl töötavad ja on avatud sisemised turud – AN puhul näiteks nii vahe- ja lõpptoodangu ning R&D puhul. Mõlemad allüksused on autonoomsed, eriti AN. Vastavalt sellele on AN ka rohkem valise turu initsiatiividelel suunatud, kui SG kasutab sisemist initsiatiivi.

Regioonile orienteeritud AK allüksuse puhul ei ole esinenud väga suurt sisemist konkurentsi – ja seega peab SGM vähem tasakaalu leidmisega tegelema. Operatsioonid, mida on üle viidud, on pigem liikunud väga selgelt jälgitavatel objektiivsetel põhjustel. Samas on AK allüksuse SGM tegev konkurentsi ja vaba jagamise vahelise tasakaalu leidmisega regiooni sees.

SG allüksuse SGM peab tasakaalu leidmiseks tegema suuremaid pingutusi. Olles sõsarfirmadega hinnakonkurentsis esineb olukordi kus koostööd võiks olla rohkem. SGMil on võimalik kasutada oma sidemeid MNC sees. SGM peab leidma ühendama organisatsiooni kui terviku (deklareeritav ühtsus) ja 'inimeste' tasandid (enda huvides tegutsemine, karjäär, töökoht).

Kui AN juht kasutab regiooni sees tasakaalu loomisel oma võimupositsiooni, siis SG puhul on tegu pigem veenmisega ja mõjutamisega (informatsiooni ja tegeliku tegutsemisvõimaluse positsioonilt). Selline tulemus ei ühti tõstatatud uurimushüpoteesiga, mis eeldas, et võimupositsiooni saab kasutada ka (all)üksustevahelise, mitte ainult üksuse (regiooni) sisese tasakaalu leidmise vahendina.

[^i]: this part should be updated with real situation
[^ii]: what about strategic business units?
[^iii]: Rat process link
[^iv]: coe link
[^vi]: should the last part of this sentence not have more ground to stand on?
[^vii]: Link this to internal competition – you get conflicting map directly linked in too
[^viii]: Last sentence should rather be replaced with a more complete review of the anatomy, physiology and psychology.
[^ix]: What was the exact distinction between functional dimension and the staff?
[^x]: Roolaht – this is not MNC exclusive
[^xi]: Am I sure there is no contradiction?
[^xii]: Link to org_culture if possible, and change back the sentence.
[^xiii]: link to initiatives at first place; also there seems to be a little contradiction – if early slack work wrote of the need for slack and GB lately wrote that the slack has been granted, then why is there still a need to dig on B statement about the need for stockiness. If ample space has been granted.
[^xiv]: It is actually added that practice market is facilitated by strong corporate culture and incentive systems that reward propensity to cooperate.